\documentclass[final]{article}

\usepackage{tikz}
\usetikzlibrary{arrows}
\usetikzlibrary{automata}
\usepackage{pgfplots}
\usepackage{amsmath}
\usepackage{graphicx}

\usepackage[margin=1in]{geometry}

\renewcommand{\inf}{\infty}

\renewcommand{\i}{\mathrm{i}}
\newcommand{\e}{\mathrm{e}}
\renewcommand{\d}{\,\mathrm{d}}
\newcommand{\diff}[2]{\frac{\mathrm{d} #1}{\mathrm{d} #2}}

\usepackage{pifont}
\usepackage{subfig}

\title{Selection of a Hele-Shaw bubble via \\ exponential asymptotics}

\author{Christopher J. Lustri$^\dagger$, \,Christopher C. Green\thanks{Department of Mathematics, Macquarie University, Sydney NSW 2109, Australia}, \, Scott W. McCue\thanks{School of Mathematical Sciences, Queensland University of Technology, Brisbane QLD 4001, Australia ({\tt scott.mccue@qut.edu.au}).}
}

\begin{document}

\newlength\figureheight
\newlength\figurewidth
\pgfplotsset{every axis y label/.append style={rotate=270}}

\maketitle

\begin{abstract}
The well-studied selection problems involving Saffman-Taylor fingers or Taylor-Saffman bubbles in a Hele-Shaw channel are prototype examples of pattern selection.  Exact solutions to the corresponding zero-surface-tension problems exist for an arbitrary finger or bubble speed, but the addition of surface tension leads to a discrete set of solution branches, all of which approach a single solution in the limit the surface tension vanishes.  In this sense, the surface tension selects a single physically meaningful solution from the continuum of zero-surface-tension solutions.  Recently, we provided numerical evidence to suggest the selection problem for a bubble propagating in an {\em unbounded} Hele-Shaw cell behaves in an analogous way to other finger and bubble problems in a Hele-Shaw channel; however, the selection of the ratio of bubble speeds to background velocity appears to follow a very different surface tension scaling to the channel cases.  Here we apply techniques in exponential asymptotics to solve the selection problem analytically, confirming the numerical results, including the predicted surface tension scaling laws.  Further, our analysis sheds light on the multiple tips in the shape of the bubbles along solution branches, which appear to be caused by switching on and off exponentially small wavelike contributions across Stokes lines in a conformally mapped plane.  These results have ramifications for exotic-shaped Saffman-Taylor fingers as well as for the time-dependent evolution of bubbles propagating in Hele-Shaw cells.

\end{abstract}

%\begin{keywords}
%velocity selection, small surface tension, Hele-Shaw cell, Laplace's equation, conformal mapping, analytic continuation, Stokes lines, asymptotics beyond all orders
%\end{keywords}
%
%\begin{AMS}
%76D27, 30E15, 35R37
%\end{AMS}

\pagestyle{myheadings} \thispagestyle{plain} \markboth{C.J. LUSTRI, C.C. GREEN \& S.W. MCCUE}{Selection of a Hele-Shaw bubble}

\section{Introduction}

Selection problems involving free boundaries in Hele-Shaw flow have received significant attention for more than 30 years since these problems are closely linked to the phenomena of viscous fingering in porous media and interfacial pattern formation for dendritic crystal growth \cite{BenJacobGarik90,Casademunt04,Homsy87,kessler88,Langer89}.  Further, the research activity in this area has driven developments in the mathematical topic of exponential asymptotics, also referred to as asymptotics beyond all orders \cite{Seguretal91}.  In this work, we continue this line of research by applying techniques in exponential asymptotics to a selection problem involving a bubble translating in a Hele-Shaw cell otherwise filled with viscous fluid.  We present new analytical results which confirm previous analytical and numerical predictions, and use Stokes cancellation arguments to explain the origin of non-convex bubbles shapes observed previously in the literature.

Of all the selection mechanisms included in Hele-Shaw models, most attention has been devoted to the role of surface tension.  In the classical Saffman-Taylor finger problem, the zero-surface-tension model has a continuum of solutions for a steadily propagating finger in a Hele-Shaw channel, where the ratio of the width of the finger to the channel  ($\lambda$, say) or, equivalently, its speed relative to the background velocity ($U$, say) is left undetermined, while experiments show that $\lambda\approx 1/2$ and $U\approx 2$ \cite{ST}.  By adding surface tension ($B$, say) to the mathematical model, both numerical studies and the application of exponential asymptotics demonstrate that for $B>0$ there is a countably infinite family of solutions with different values of $\lambda$ and $U$.  In the limit $B\rightarrow 0$, a single zero-surface-tension solution is selected from the continuum with $\lambda=1/2$ and $U=2$ \cite{Chapman99,CombescotEtAl,GardinerEtAlb,Hong,McLeanSaffman,Shraiman,Tanveer872,VDB1983}.

In the present study, we are focussed on the selection problem for a single steadily moving bubble.  When the bubble is propagating in a channel geometry, the analysis is analogous to the finger problem just mentioned.  Again, the zero-surface-tension model has exact solutions with a continuum of velocities $U$ \cite{TS}; however, the model with surface tension provides a countably infinite family of solutions, each with a different value of $U$ for a fixed $B>0$, and taking the limit $B\rightarrow 0$ selects a single zero-surface-tension solution with $U=2$ \cite{Combescot,Tanveer86,Tanveer87,Tanveer89}.  It is important for our study to note that for both the finger and bubble problem in a Hele-Shaw channel, the small-surface-tension scaling is known to be $U\sim 2- k B^{2/3}$, where $k$ is a constant that depends on the solution branch number.

Very recently, we revisited this selection problem, except we concentrated on the case in which the Hele-Shaw cell was unbounded \cite{GreenLustriMcCue}.  By using a combination of conformal mapping and numerical methods, we provided evidence to suggest that the bubble selection problem in an unbounded Hele-Shaw cell behaves qualitatively like in the channel case, however the small-surface-tension scaling appeared to be $U\sim 2- k B^{2}$, where again $k$ is a constant which is different for each solution branch.  Since that publication, we have learned that in fact Hong \& Family~\cite{HongFamily88} made the same prediction.
Here, we use exponential asymptotics to confirm this scaling analytically, and go further by showing that
\begin{equation}
U\sim 2-\frac{\pi^2}{2^{13}}\left(\frac{\Gamma(1/4)}{\Gamma(7/4)}\right)^4(2m+1)^4\,B^2
\quad\mbox{as}\quad B\rightarrow 0,
\label{eq:mainresult}
\end{equation}
where the branch number $m$ takes values $m=1,2,\dots$.  This apparent contradiction when compared to the channel geometry is possible because, for the channel problem, the limit of increasing the channel width to infinity does not commute with the vanishing surface tension limit, thus it turns out the channel geometry is not appropriate in the small bubble limit \cite{GreenLustriMcCue,HongFamily88,Tanveer89}.

In order to derive (\ref{eq:mainresult}), we employ a combination of three broad methodologies.  First, we reformulate the problem using the same type of conformal mapping as Tanveer~\cite{Tanveer86,Tanveer872,Tanveer89}, for example (who applied it to both bubble and finger problems).  This approach involves mapping from unit disc to the physical plane, with points on the unit circle mapping to the bubble boundary.  Second, the techniques in exponential asympotics that we apply are based on ideas set out by Chapman, King \& Adams for nonlinear ordinary differential equations, where the (divergent) asymptotic series is truncated optimally and the exponentially small remainder terms are observed to `switch on' across Stokes lines~\cite{chapman98}.  These Stokes lines emerge from singularities in the analytical continuation of the leading order term and typically intersect the part of the complex plane corresponding to the physical problem of interest (which is the real axis in many problems, but the unit circle in our problem).  Note that this method has been successfully applied to a variety of problems in fluid mechanics, including two- and three-dimensional water waves \cite{chapmanVDB2,lustrichapman13,lustrichapman14,lustrimccue12,lustrimccue13,trinhchapman13,trinhchapman14}. Finally, the third key idea is that of selection.  Our analysis will produce what appears to be exponentially larger contributions in certain regions of the unit circle (corresponding to the bubble boundary).  In order to avoid this unphysical scenario, we force contributions from two singularities to cancel each other, thus deriving a solvability condition which effectively leads to (\ref{eq:mainresult}).  This idea of cancelling exponentially large terms in order to determine a solvability condition is similar in theme to that used by Chapman~\cite{Chapman99,ChapmanKing} for other selection problems involving Hele-Shaw fingers.  However, the mapping is completely different in those cases, and so the details are not the same.

Apart from deriving the scaling law (\ref{eq:mainresult}), the focus of the present study is to highlight the shape of the bubbles as surface tension increases along each solution branch.  For small surface tension, these bubbles are all almost circular, but as the surface tension parameter $B$ increases, the bubbles become non-convex with a number of tips or dimples, depending on the branch number \cite{GreenLustriMcCue}.  This wakelike behaviour has also been noted by Tanveer~\cite{Tanveer87} for a bubble in a channel geometry and is analogous to that observed at the front of a Saffman-Taylor finger \cite{GardinerEtAlb}.  By analysing the remainder term along the unit circle between the Stokes intersection points, we explain the birth of these oscillations, and demonstrate analytically that the solutions on the $m$th branch develop $m+1$ tips (as observed numerically in \cite{GreenLustriMcCue}).

The structure of our paper is as follows.  In Section~\ref{sec:formulation}, we present the dimensionless problem, the conformal mapping and the exact solutions for zero surface tension.   In Section~\ref{sec:seriesexpansion} we write out a (divergent) power-series expansion in powers of $B$ and determine analytical results for early and late-order terms.  By truncating our series optimally, in Section~\ref{sec:exponential} we study how exponentially small terms are switched on across Stokes lines, ultimately leading to a solvability condition for non-zero-surface-tension solutions, which we derive in Section~\ref{S:Stokes Structure}.  This solvability condition implies the scaling (\ref{eq:mainresult}), which we compare favourably with the numerical results from \cite{GreenLustriMcCue}.  Section~\ref{sec:bubbleshape} is devoted to our explanation for the exotic non-convex bubbles shapes for sufficient large surface tension.  Finally, in Section~\ref{sec:discussion} we summarise the interesting features of our study and discuss the important of the main results.  In this section we highlight they key challenges involved in applying exponential asymptotics to our selection problem.  We comment on the significance of (\ref{eq:mainresult}) and how it relates to the very different scalings appropriate for Hele-Shaw channels.  Further, we discuss the connections between the (double-tipped) non-convex bubble shapes identified in our study and numerical/experimental findings from bubbles propagating in standard and non-standard Hele-Shaw cells.

%Finally, we identify some open problems, including the stability of the bubble solutions and some more complicated selection problems involving multiple bubbles.

%%%%%%%%% CCG putting in stuff

\section{Formulation and zero-surface-tension solution}\label{sec:formulation}

\subsection{Governing equations in the physical plane}

We first summarise the problem formulation presented in \cite{GreenLustriMcCue} for a single bubble steadily translating with speed $U$ in an unbounded Hele-Shaw cell. Let $D$ be the unbounded two-dimensional region in the complex $z$-plane containing incompressible fluid exterior to the bubble, which is assumed to be vertically symmetric. The system of interest describes the velocity potential $\phi$ and the streamfunction $\psi$ in a reference frame co-travelling with the bubble. After the introduction of non-dimensional variables, this system is \cite{GreenLustriMcCue}
\begin{align}
\nabla^2 \phi &= 0, \quad z \in D,\\
\phi + U x &= B x + \phi_0, \quad z \in \partial D,\\
\psi &= 0, \quad z\in \partial D,\label{0.kinematic}\\
\phi &\sim (1-U)x, \quad |z| \rightarrow \infty.
\end{align}
Here, $B$ is a non-dimensional surface tension parameter, $\phi_0$ is a real constant (whose value is arbitrary), and $1 < U \leq 2$.

\subsection{Conformal mapping}

\begin{figure}
\centering
\includegraphics[width=0.8\textwidth]{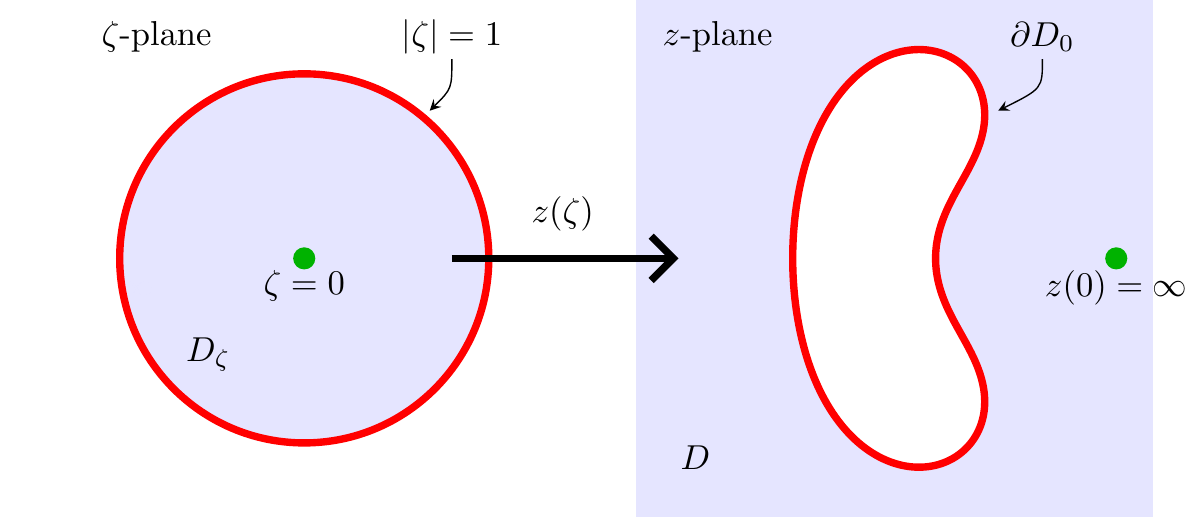}
%\begin{tikzpicture}
%[xscale=1.5,>=stealth,yscale=1.5]
%
%\node[white] at (-5.5,0) [below] {$z(0)=\infty$};
%\fill[blue,opacity=0.1] (-4,0) circle (1.25);
%\draw[line width=0.75mm,red] (-4,0) circle (1.25);
%
%\fill[blue,opacity=0.1] (-1.75,-1.75) -- (1.75,-1.75) -- (1.75,1.75) -- (-1.75,1.75) -- cycle;
%\fill[white] plot[smooth] file {SchematicBubble.txt};
%\draw[line width=0.75mm,red] plot[smooth] file {SchematicBubble.txt};
%
%\node at (-4.65,-0.65) {$D_{\zeta}$};
%\fill[black!30!green] (-4,0) circle (0.075);
%\node at (-4,0) [below] {$\zeta=0$};
%\node at (-3,1.5) {$|\zeta| = 1$};
%\draw [->] (-3,1.35) .. controls (-3,1.15)  .. (-3.15,1);
%
%\draw[line width=0.75mm] (-3,0) -- (-1.5,0);
%\draw[line width=0.75mm] (-1.65,0.15) -- (-1.5,0) -- (-1.65,-0.15);
%\fill[black!30!green] (1.5,0) circle (0.075);
%\node at (1.5,0) [below] {$z(0)=\infty$};
%
%\node at (1,1.5) {$\partial D_0$};
%\draw [->] (1,1.35) .. controls (1,1.15)  .. (0.7,1);
%
%\node at (-2.25,0.1) [above] {$z(\zeta)$};
%
%\node at (-1.35,-1.35) {$D$};
%\node at (-5,1.5) {$\zeta$-plane};
%\node at (-1.2,1.5) {$z$-plane};
%
%\end{tikzpicture}
\caption{Schematic of (a) the pre-image $\zeta$-plane being the interior of the unit circle, and (b) the target image domain in the physical $z$-plane, exterior to the bubble. This mapping permits us to study the problem in a fixed domain in the $\zeta$-plane, and use this to determine the location of the bubble boundary in the physical plane.}
\label{F:Bubble_Setup}
\end{figure}

The complex potential function $w(z) = \phi + \i \psi$, which is analytic and single-valued everywhere in $D$, is sought by considering the composition $W(\zeta)=w(z(\zeta))$, where $z(\zeta)$ is a conformal map which transplants the interior $D_\zeta$ of the unit circle in a complex $\zeta$-plane to the exterior of the bubble in the $z$-plane (see Figure \ref{F:Bubble_Setup}).

Re-writing the boundary conditions (2.2)--(2.4) in terms of the complex variable $\zeta$ results in:
\begin{align}
\mathrm{Re}[W(\zeta)+Uz(\zeta)] &= B\kappa + \mathrm{constant}, \quad \mbox{on} \quad |\zeta|=1,\label{0.bc1}\\
\mathrm{Im}[W(\zeta)] &= 0, \quad \mbox{on} \quad |\zeta|=1,\\
W(\zeta) &\sim a(1-U)/\zeta, \quad \zeta \rightarrow 0.
\end{align}
The kinematic boundary condition \eqref{0.kinematic} becomes $\mathrm{Im}[W(\zeta)]=0$, on $|\zeta|=1$, and the far-field condition (2.4) becomes $W(\zeta) \sim a(1-U) / \zeta$, $\zeta \rightarrow 0$.

We introduce $W_0(\zeta)$ and $z_0(\zeta)$ as the complex potential and conformal map, respectively, for a zero-surface-tension bubble in this system. This pair of solutions is \cite{Crowdy2009,TS}
\begin{equation*}
W_0(\zeta) = a(1-U)\left(\zeta+\frac{1}{\zeta}\right),\qquad  z_0(\zeta) = \frac{a}{\zeta} + a\left(1-\frac{2}{U}\right)\zeta.
\end{equation*}
Here, $a$ is a real constant related to the bubble area. Fixing the area of the bubble to be $\pi$, without loss of generality, results in $a = U/2\sqrt{U-1}$ when $B=0$.

The family of non-zero-surface-tension solutions is given by
\begin{equation*}
W(\zeta) \equiv W_0(\zeta), \qquad z(\zeta)=z_0(\zeta)+f(\zeta).
\end{equation*}
Thus, for a fixed value of the surface tension parameter $B$, the problem is to solve for $f(\zeta)$ and the real numbers $U$ and $a$, constraining the bubble area to be $\pi$, and enforcing the boundary condition \eqref{0.bc1}, which gives
\begin{equation}
\label{0:Geq}
U \mathrm{Re}[f(\zeta)] = -  B \frac{\Bigg(1 + \mathrm{Re}\Bigg[\dfrac{\zeta(z''_0(\zeta) + f''(\zeta))}{z_0'(\zeta) + f'(\zeta)}\Bigg]\Bigg)}{|z'_0(\zeta) + f'(\zeta)|} \quad \mbox{on} \quad |\zeta|=1.
\end{equation}
Here, we have used a formula for the signed curvature $\kappa$ written in terms of the conformal map $z(\zeta)$.

Trivial solutions for any $B \ne 0$ are such that
\begin{equation*}
f(\zeta)=\tfrac{B}{2}, \quad U=2, \quad a=1,
\end{equation*}
corresponding to circular bubbles of unit radius. We call these solutions the $m=0$ branch. For nontrivial solutions, a numerical scheme was developed in \cite{GreenLustriMcCue} which involved writing out $f$ as a Taylor series in $\zeta$, truncating after a finite of terms, and applying (\ref{0:Geq}) at equally spaced points along the unit circle $|\zeta|=1$.  For a fixed $B>0$, this numerical scheme computed a number of solutions along different branches.  As mentioned in the Introduction, the numerical results in \cite{GreenLustriMcCue} included the predicted scaling of these solution branches $U\sim 2- k B^{2}$ as $B\rightarrow 0$ as well as examples of exotic bubble shapes along each branch as $B$ increases.

%%%%%%%%% CJL putting in stuff

\section{Power series expansion}\label{sec:seriesexpansion}

We are concerned with analysing (\ref{0:Geq}) in the singular limit $B\rightarrow 0$.  With this in mind, we begin with the power series expansion
\begin{equation}\label{0:fseries}
f(\zeta) \sim \sum_{n=1}^{\inf} B^n f_n(\zeta)
\quad\mbox{as}\quad B\rightarrow 0.
\end{equation}
Recall that the shape of the bubble is given by the mapping function $z(\zeta)=z_0(\zeta)+f(\zeta)$, where $z_0$ is the zero-surface-tension solution.  Therefore, the first terms in this series, $Bf_1$, $B^2f_2$, etc., provide correction terms to $z_0$ in the limit $B\ll 1$.

\subsection{Series expansion}

Applying (\ref{0:fseries}) to \eqref{0:Geq}, and matching to first order in the limit $B \rightarrow 0$, we obtain
\begin{equation}
\mathrm{Re}[f_1(\zeta)] = \frac{4(U-1)}{aU^3(1 + q^2\zeta^2)^{3/2}(1 + q^2/\zeta^2)^{3/2}} = h(\zeta)
\quad \mbox{on} \quad |\zeta|=1,
\label{eq:goveqnf1}
\end{equation}
where $q^2 = 2/U - 1$. The choice of sign in $q$ is arbitrary, and we will select the positive branch of the square root in all subsequent analysis, so that $q = \sqrt{2/U-1}$.
%\begin{equation*}
%q=\sqrt{\tfrac{2}{U}-1}.
%\end{equation*}

Applying the Schwarz integral formula to \eqref{eq:goveqnf1}, we find that
\begin{equation*}
f_1(\zeta) = \frac{1}{2\pi\i} \oint_{|\zeta'| = 1}h(\zeta') \frac{\zeta' + \zeta}{\zeta' - \zeta}\frac{\d \zeta'}{\zeta'} = I_1(\zeta),\qquad |\zeta| < 1.
\end{equation*}
This integral is analytic everywhere on the unit disc, despite $h(\zeta)$ being singular at $\zeta=\pm\i q$.  Analytic continuation of $f_1(\zeta)$ \textit{outside} the unit circle
gives
\begin{equation*}%\label{0:f1}
f_1(\zeta) = I_1(\zeta) + 2h(\zeta),\qquad |\zeta| >1.
\end{equation*}
Consequently, $f_1(\zeta)$ contains singularities at $\zeta = \pm \i/q$. From this expression, we find
\begin{equation}\label{0:f1local}
f_1(\zeta) \sim \frac{\sqrt{\pm\i q}\, U}{4 \sqrt{2} a (U-2)\sqrt{U-1}(\zeta\mp \i/q)^{3/2}} \qquad \mathrm{as} \qquad \zeta \rightarrow \pm\i/q,
\end{equation}
where the upper and lower signs correspond in each case.

In order to find a recursion relation for subsequent terms of the series, we analytically continue the governing equation \eqref{0:Geq} off the unit circle, obtaining
\begin{equation}\label{0:analytic}
\frac{U}{2} [f(\zeta) + \overline{f}(1/\zeta)] =    -B\frac{\Bigg(1 + \dfrac{1}{2}\Bigg[\dfrac{\zeta(z''_0(\zeta) + f''(\zeta))}{z_0'(\zeta) + f'(\zeta)} + \dfrac{(\overline{z}''_0(1/\zeta) + \overline{f}''(1/\zeta))}{\zeta(\overline{z}_0'(1/\zeta) + \overline{f}'(1/\zeta))}\Bigg]\Bigg)}{(z'_0(\zeta) + f'(\zeta))^{1/2}(\overline{z}'_0(1/\zeta) + \overline{f}'(1/\zeta))^{1/2}}  ,
\end{equation}
where $\overline{f}(\zeta)$ is shorthand for the analytic expression $\overline{f(\overline{\zeta})}$, in which the bar represents complex conjugation. This expression, valid for all complex $\zeta$, reduces to \eqref{0:Geq} on the unit circle $|\zeta|=1$.

Applying \eqref{0:fseries} gives a recursion relation for $f_n$,
\begin{align}
\frac{U}{2} [f_n(\zeta) + \overline{f}_n(1/\zeta)] = -&\frac{\zeta f_{n-1}''(\zeta)}{2(z'_0(\zeta) )^{3/2}(\overline{z}'_0(1/\zeta))^{1/2}} -\frac{ \overline{f}_{n-1}''(1/\zeta)}{2\zeta(z'_0(\zeta) )^{1/2}(\overline{z}'_0(1/\zeta))^{3/2}} \nonumber \\
& + \left[\frac{2\zeta\overline{z}_0'(1/\zeta) z_0'(\zeta) + \overline{z}_0''(1/\zeta) z_0'(\zeta) + 3 \zeta^2 \overline{z}'_0(1/\zeta) z_0''(\zeta)}{4\zeta (z'_0(\zeta) )^{5/2}(\overline{z}'_0(1/\zeta))^{3/2}}\right]f_{n-1}'(\zeta) \nonumber \\
& + \left[\frac{2\zeta\overline{z}_0'(1/\zeta) z_0'(\zeta) + 3\overline{z}_0''(1/\zeta) z_0'(\zeta) +  \zeta^2 \overline{z}'_0(1/\zeta) z_0''(\zeta)}{4\zeta (z'_0(\zeta) )^{3/2}(\overline{z}'_0(1/\zeta))^{5/2}}\right]\overline{f}_{n-1}'(1/\zeta) + \ldots.\label{0:recursion0}
\end{align}
Given the solution for $f_1$, we can, in principle, apply \eqref{0:recursion0} to determine $f_2$, then $f_3$, and so on.  These calculations are algebraically complicated and, given these extra terms are not required in the subsequent analysis, we do not pursue them.  We shall, however, use \eqref{0:recursion0} to determine the form of the late-order terms $n\gg 1$ which are required for our exponential asymptotics.  For this purpose it turns out that the omitted terms in \eqref{0:recursion0} will be subdominant compared to those that have been retained in the limit $n\rightarrow\infty$.

\subsection{Late-order terms}

\subsubsection{Factorial over power divergence}

We see from \eqref{0:recursion0} that calculating $f_n$ requires taking two derivatives of $f_{n-1}$. Consequently, we expect that any singularities in $f_{n-1}$ of strength $p$ will correspond to singularities in $f_n$ with strength $p+2$. As a consequence of this repeated differentiation, we expect that the terms in the series \eqref{0:fseries} will diverge in a  factorial-over-power fashion, as discussed in \cite{Dingle}. We exploit this predictable behaviour in order to determine the asymptotic behaviour of $f_n(\zeta)$ in the limit that $n \rightarrow \inf$, using a late-order term ansatz, as described in Chapman, King \& Adams \cite{chapman98}.

Recalling that we require two derivatives of $f_{n-1}$ to obtain $f_n$, we set the late-order terms to be a sum of terms with the form
\begin{equation}\label{0:ansatz}
f_n(\zeta) \sim \frac{F(\zeta)\Gamma(2n + \gamma)}{\chi(\zeta)^{2n + \gamma}}\qquad \mathrm{as} \qquad n \rightarrow \inf,
\end{equation}
where $F$, $\gamma$ and $\chi$ are independent of $n$, and $\chi(\zeta) = 0$ at the early-order singularities of $f_1$, located at $\zeta = \pm \i/q$. The late-order terms also contain contributions from $\overline{f}_n(1/\zeta)$, which is singular at $\zeta = \pm\i q$. From \cite{Dingle} we know that these terms may be considered separately to the $f_n$ terms. We will perform a detailed analysis for the $f_n(\zeta)$ terms described by \eqref{0:ansatz}, and add the corresponding results for $\overline{f}_n(1/\zeta)$ upon completing the analysis.

\subsubsection{Determining the singulant $\chi$} Applying the late-order ansatz \eqref{0:ansatz} to the recursion relation \eqref{0:recursion0} gives
\begin{equation*}%\label{0:recursion}
\frac{U F(\zeta)\Gamma(2n + \gamma)}{2 \chi(\zeta)^{2n + \gamma}}  = -\frac{\zeta}{2 (z_0'(\zeta))^{3/2}(\overline{z}'_0(1/\zeta))^{1/2}}\frac{(-\chi'(\zeta))^2 F(\zeta)\Gamma(2n + \gamma)}{\chi(\zeta)^{2n + \gamma}} + \ldots,
\end{equation*}
where the omitted terms are $o(f_n)$ as $n\rightarrow \inf$. Matching to leading order in this limit gives 
\begin{equation*}
U=-\frac{\zeta (\chi'(\zeta))^2}{(z_0'(\zeta))^{3/2}(\overline{z}'_0(1/\zeta))^{1/2}},
\end{equation*} 
which implies that
\begin{equation}
\label{eqn:chidash}
\chi'(\zeta)=+\mathrm{i}\sqrt{U}\,
\frac{(z_0'(\zeta))^{3/4}(\overline{z}'_0(1/\zeta))^{1/4}}{\zeta^{1/2}}.
\end{equation}
The positive sign in \eqref{eqn:chidash} is chosen because a full exponential asymptotic analysis shows that only the positive choice of sign produces exponentially small contributions to the bubble shape.  

There are two singulants of interest, one of which vanishes at the singularity $\zeta = \i /q$ and the other which vanishes at $\zeta = -\i /q$.  We refer to these as $\chi_U$ and $\chi_L$, respectively.
By integrating \eqref{eqn:chidash}, these singulants are given by
\begin{equation}
\chi_{U}(\zeta) =  \int_{\i/q}^{\zeta} \chi'(s) \d s,\quad
\chi_{L}(\zeta) =  \int_{-\i/q}^{\zeta} \chi'(s) \d s.\label{0:singulants}
\end{equation}
As these integrals are too challenging to perform analytically, we evaluate the two singulants numerically using contour integration.

\begin{figure}[tb]
\centering
\includegraphics[width=0.8\textwidth]{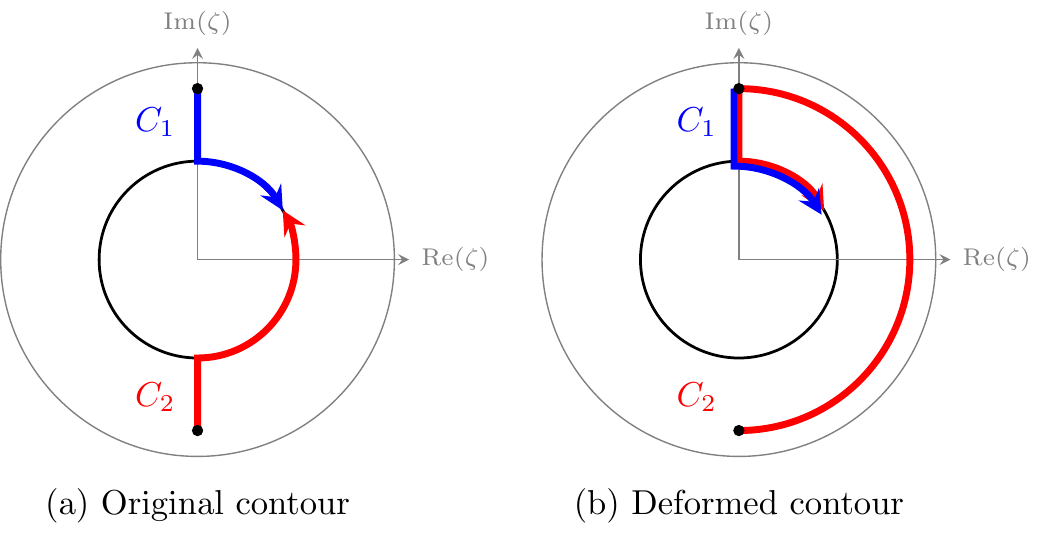}
%\begin{tikzpicture}
%[xscale=1.,>=stealth,yscale=1.]
%
%\draw[white] (0,0) circle (2);
%\draw[thick] (0,0) circle (1);
%
%%\draw[gray] (-3,3) -- (3,3) -- (3,-3) -- (-3,-3) -- cycle;
%\draw[gray] (0,0) circle (2);
%\draw[gray,->] (0,0) -- (2.15,0) node [right] {\scriptsize{$\mathrm{Re}(\zeta)$}};
%\draw[gray,->] (0,0) -- (0,2.15) node [above] {\scriptsize{$\mathrm{Im}(\zeta)$}};
%
%\draw[line width=2pt,blue,->] (0,1.737) -- (0,1) arc (90:30:1);
%\draw[line width=2pt,red,->] (0,-1.737) -- (0,-1) arc (-90:30:1);
%\node[blue] at (-0.1,1.4) [left] {$C_1$};
%\node[red] at (-0.1,-1.4) [left] {$C_2$};
%
%\filldraw[] (0,1.737) circle (0.05);
%\filldraw[] (0,-1.737) circle (0.05);
%
%\node at (0,-2.5) {(a) Original contour};
%\node at (5.5,-2.5) {(b) Deformed contour};
%
%\draw[white] (5.5,0) circle (2);
%\draw[thick] (5.5,0) circle (1);
%
%%\draw[gray] (-3,3) -- (3,3) -- (3,-3) -- (-3,-3) -- cycle;
%\draw[gray] (5.5,0) circle (2);
%\draw[gray,->] (5.5,0) -- (7.65,0) node [right] {\scriptsize{$\mathrm{Re}(\zeta)$}};
%\draw[gray,->] (5.5,0) -- (5.5,2.15) node [above] {\scriptsize{$\mathrm{Im}(\zeta)$}};
%
%\draw[line width=2pt,red,->] (5.5,-1.737) arc (-90:90:1.737) -- (5.5,1) arc (90:30:1);
%\draw[line width=2pt,blue,->] (5.45,1.737) -- (5.45,0.95) arc (90:32:1.05);
%\node[blue] at (5.4,1.4) [left] {$C_1$};
%\node[red] at (5.4,-1.4) [left] {$C_2$};
%
%\filldraw[] (5.5,1.737) circle (0.05);
%\filldraw[] (5.5,-1.737) circle (0.05);
%
%\end{tikzpicture}
\caption{(a) Contours of integration for the singulant integrals in \eqref{0:singulants}. Each contour begins at the associated singularity (the blue contour for $\chi_U$ begins at $\zeta=\i/q$, while the red contour for $\chi_L$ begins at $\zeta=-\i/q$) and terminates at a point $\zeta$ on the bubble boundary. (b) The contour for $\chi_L$ is deformed so that it passes through the upper singularity, leading to (\ref{4:intrelation}).}
\label{Fig:contour}
\end{figure}

Clearly, since $\chi'_U=\chi'_L$ (both given by \eqref{eqn:chidash}), the two singulants $\chi_U$ and $\chi_L$ differ by a constant only.  We determine this constant via deforming the contour of integration for $\chi_L$ so that it passes through $\zeta = \i/q$, as illustrated in Figure \ref{Fig:contour}.  As a result, we find
\begin{equation}
\chi_{L}(\zeta) = \int_{-\i/q}^{\i/q} \chi'(s)\d s + \chi_U(\zeta).\label{4:intrelation}
\end{equation}
%\begin{align}
%\chi_{L}(\zeta) &=  \i\sqrt{U} \int_{-\i/q}^{\i/q} \frac{(z_0(s))^{3/4}(\overline{z}'_0(1/s))^{1/4}}{s^{1/2}} \d s +  \i\sqrt{U} \int_{\i/q}^{\zeta} \frac{(z_0(s))^{3/4}(\overline{z}'_0(1/s))^{1/4}}{s^{1/2}} \d s,\nonumber\\
%&=  \i\sqrt{U} \int_{-\i/q}^{\i/q} \frac{(z_0(s))^{3/4}(\overline{z}'_0(1/s))^{1/4}}{s^{1/2}} \d s + \chi_U(\zeta).\label{4:intrelation}
%\end{align}
The relationship \eqref{4:intrelation} turns out to be crucial for our subsequent analysis in Section~\ref{S:Stokes Structure}, where we apply Stokes switching arguments to derive a discrete set of possible values for $\chi_L-\chi_U$.  Combining that result with \eqref{4:intrelation} will produce the solvability condition we aim to determine.

Due to the conjugate symmetry of $z_0(\zeta)$, the integrand $\chi'(\zeta)$ also has conjugate symmetry. Hence, the integral of $\chi'(\zeta)$ from $-\i/q$ to $\i/q$ must be real valued and, furthermore, from \eqref{4:intrelation} we note that
%\begin{align}
%\chi_{L}(1) &=  \int_{-\i/q}^{\i/q} \chi'(s) \d s +  \chi_U(1),\nonumber\\
%&=  2\int_{-\i/q}^{1} \chi'(s) \d s +  \chi_U(1),\nonumber\\
%&=  2\chi_L(1)+\chi_U(1).
%\end{align}
\begin{equation}
\int_{-\i/q}^{\i/q} \chi'(s) \d s=-2\chi_U(1)=2\chi_L(1).
\label{eq:chiU1chiL1}
\end{equation}
This relationship will be used later in Section~\ref{sec:bubbleshape} to explain the multi-tipped nature of the bubble shapes along solution branches as $B$ increases.

\subsubsection{Determining the prefactor $F$}
By again applying the late-order ansatz \eqref{0:ansatz} to the recursion relation \eqref{0:recursion0}, we can match at $\mathcal{O}(f_{n-1})$ to obtain an equation for the prefactor $F(\zeta)$, given by
\begin{equation*}%\label{0:prefactor0}
\frac{F'(\zeta)}{F(\zeta)} = \frac{1}{4}\left(\frac{2}{\zeta} - \frac{2\chi''(\zeta)}{\chi'(\zeta)} + \frac{\overline{z}_0''(1/\zeta)}{\zeta^2 \overline{z}_0'(1/\zeta)} + \frac{3 z_0''(\zeta)}{z_0'(\zeta)}\right).
\end{equation*}
Note that $\chi'_U = \chi'_L$, and therefore the singulant derivatives in this expression are independent of the choice of singulant. Integrating both sides gives
\begin{equation*}%\label{0:prefactor1}
\log(F(\zeta)) = \tfrac{1}{2}\log(\zeta) - \tfrac{1}{4}\log(2-U+U\zeta^2) - \tfrac{1}{2}\log(\chi'(\zeta)) + \tfrac{3}{4}\log(z_0'(\zeta)) + c,
\end{equation*}
where $c$ is an arbitrary constant. This can be rewritten as
\begin{equation}\label{0:prefactor}
F(\zeta)= \Lambda \left(\frac{U + 2\zeta^2 - U\zeta^2}{2-U+U\zeta^2}\right)^{3/8},
\end{equation}
where $\Lambda = \mathrm{e}^c$, and is therefore also an arbitrary constant.

We find by direct calculation that, as $\zeta \rightarrow \i/q$,
\begin{gather}
z_0(\zeta) \sim -2\i a q, \qquad z_0'(\zeta) \sim 2\i a q^{3/2} (\zeta - \i/q),\qquad  z_0''(\zeta) \sim 2\i a q^{3/2},\nonumber\\
\overline{z_0}'(1/\zeta) \sim \frac{4 a (U-1)}{U (2-U)},\qquad  \overline{z_0}''(1/\zeta) \sim -2\i a q^{-3/2}.\label{0:localvals}
\end{gather}
We can use this information to determine the local behaviour of $\chi$ near the singular point at $\zeta = \i/q$. We assume $\chi$ has the asymptotic behaviour $\chi \sim \alpha (\zeta - \i/q)^{\beta}$ and apply \eqref{0:localvals} to \eqref{0:singulants} in the limit that $\zeta \rightarrow \i/q$, to give
\begin{align*}%\label{0:innersingulant}
\alpha =  \pm\tfrac{8a}{7}(1+\i)  [2 \i (1-U) q^9]^{1/4},\qquad \beta = \tfrac{7}{4}.% \qquad k = \tfrac{3}{8},
\end{align*}
Comparing this expression with $\chi_U$ shows that the positive choice of sign in $\alpha$ gives the appropriate local singulant behaviour.

Furthermore, it is straightforward to show from \eqref{0:prefactor} that, as $\zeta \rightarrow \i/q$,
\begin{equation*}
F(\zeta) \sim \Lambda C (\zeta - \i/q)^{3/8} \qquad C = \left(\frac{\sqrt{U}(U-2)^{3/2}}{2(U-1)}\right)^{3/8}.
\end{equation*}

\subsubsection{Determining the parameter $\gamma$}
We now have sufficient information to determine $\gamma$ in \eqref{0:ansatz}. In the neighbourhood of the singularity $\zeta = \i/q$, we find that the late-order terms behave like
\begin{equation}\label{0:innerofouter}
f_n(\zeta) \sim \frac{\Lambda C (\zeta-\i/q)^{3/8}\Gamma(2n+\gamma)}{[\alpha (\zeta - \i/q)^{7/4}]^{2n + \gamma}}, \qquad \mathrm{as} \qquad n\rightarrow \inf,\quad\zeta \rightarrow \i/q.
\end{equation}
We recall that the local behaviour of $f_1(\zeta)$, given in \eqref{0:f1local}, has a singularity at $\zeta = \i/q$ of strength $3/2$. To ensure that this is consistent with the late-order ansatz, we require that $3/8-7(2+\gamma)/4 = -3/2$, which gives $\gamma = -13/14$.

\subsubsection{Determining the constant $\Lambda$}
In order to determine $\Lambda$, we match the behaviour of the late-order terms with the behaviour of the solution in the neighbourhood of the singularity at $\zeta = \i/q$, obtained numerically. The analysis is performed in Appendix \ref{App:Lambda}, with an example numerical value obtained for the case $U = 1.5$, giving $\Lambda \approx -0.079 + 0.347\i$. The numerical process can easily be repeated in order to determine the value of $\Lambda$ for any choice of $U$.

\section{Exponential asymptotics}\label{sec:exponential}

\subsection{Optimal truncation}

We truncate the divergent asymptotic series after $N-1$ terms, to obtain
\begin{equation}\label{0:fseries_truncated}
f(\zeta) = \sum_{j=1}^{N-1} B^j f_j(\zeta) + R_N(\zeta),
\end{equation}
where $R_N(\zeta)$ is the remainder term. Berry \cite{BerryHowls90} showed that if the truncation point is chosen optimally, this remainder will be exponentially small. Typically, the optimal truncation point may be found using a heuristic described in \cite{Boyd99}, in which the series \eqref{0:fseries} is truncated after the smallest term.  Approximating series terms using the ansatz \eqref{0:ansatz} and applying this strategy gives $N\sim |\chi|/2\sqrt{B}$ in the limit that $B \rightarrow 0$ and $N \rightarrow \inf$. We therefore set $N = |\chi|/2\sqrt{B} + \omega$, where $\omega$ is a constant in the range $0 \leq \omega < 1$ chosen to ensure that $N$ takes an integer value.

Substituting the truncated series \eqref{0:fseries_truncated} into the governing equation \eqref{0:analytic}, and simplifying using the recursion relation \eqref{0:recursion0} to eliminate terms, we eventually obtain
\begin{align}\label{0:RNeq}
U R_N - U  B^N f_N = -& \frac{B \zeta R_N''(\zeta)}{(z'_0(\zeta) )^{3/2}(\overline{z}'_0(1/\zeta))^{1/2}} \nonumber \\
& + B\left[\frac{2\zeta\overline{z}_0'(1/\zeta) z_0'(\zeta) + \overline{z}_0''(1/\zeta) z_0'(\zeta) + 3 \zeta^2 \overline{z}'_0(1/\zeta) z_0''(\zeta)}{2\zeta (z'_0(\zeta) )^{5/2}(\overline{z}'_0(1/\zeta))^{3/2}}\right]R_N'(\zeta) + \ldots.
\end{align}
Away from the Stokes line, the inhomogeneous term in \eqref{0:RNeq} will be smaller than the remaining terms in the limit that $B \rightarrow 0$. Consequently, we can determine the asymptotic behaviour of $R_N(\zeta)$ away from the Stokes line by applying the WKB (or Green-Liouville) method.

\subsection{Stokes switching}

A straightforward WKB ansatz suggests that $R_N \sim A F(\zeta) \e^{-\chi(\zeta)/\sqrt{B}}$, where $A$ is some constant.  In order to capture the Stokes switching in the neighbourhood of the Stokes line, we write
\begin{equation*}
R_N \sim A(\zeta) F(\zeta) \e^{-\chi(\zeta)/\sqrt{B}} \quad \mathrm{as} \quad B \rightarrow 0,
\end{equation*}
where $A$ is a now a Stokes multiplier that is essentially constant away from the Stokes line, but varies rapidly as the Stokes line is crossed.

Using the singulant equation \eqref{0:singulants} and the prefactor equation \eqref{0:prefactor}, we are able to cancel a number of terms, giving to leading order as $B \rightarrow 0$,
%\begin{equation}
%-\frac{\zeta \chi'(\zeta) B^{1/2}}{U (z'_0(\zeta) )^{3/2}(\overline{z}'_0(1/\zeta))^{1/2}} A'(\zeta) F(\zeta)\e^{-\chi(\zeta)/\sqrt{B}} \sim  B^{N-1/2} f_N,
%\end{equation}
%which reduces to
\begin{equation*}
 A'(\zeta) F(\zeta)\e^{-\chi(\zeta)/\sqrt{B}} \sim  \chi'(\zeta)B^{N-1/2} f_N.
\end{equation*}
Recalling that $N$ is large as $B \rightarrow 0$, we apply the late-order ansatz to obtain
\begin{equation}\label{0:smoothing1}
 A'\sim \frac{\chi' B^{N-1/2}\Gamma(2N-13/14) }{\chi^{2N-13/14}}\,\e^{\chi/\sqrt{B}}.
\end{equation}
By applying the chain rule, we write the singulant as a function of $\chi$,  and express the independent variable in terms of polar coordinates so that $\chi = \rho \e^{\i\theta}$. Taking the derivative in the radial direction, this transformation gives
\begin{equation}
\diff{}{\zeta} = -\frac{\i\chi' \e^{-\i\theta}}{\rho}\diff{}{\theta},
\end{equation}
while the optimal truncation point becomes $N = \rho/2\sqrt{B} + \omega.$ Hence \eqref{0:smoothing1} becomes
\begin{equation*}%\label{0:smoothing2}
\diff{A}{\theta} \sim \i\rho\e^{\i\theta}B^{\rho/2\sqrt{B} + \omega-1/2} \exp\left(\frac{\rho\e^{\i\theta}}{\sqrt{B}}\right) \frac{\Gamma(\rho/\sqrt{B} + 2\omega-13/14)}{(\rho\e^{\i\theta})^{\rho/\sqrt{B} + 2\omega-13/14}}.
\end{equation*}
in the $B \rightarrow 0$ limit. We can now apply Stirling's formula \cite{abramstegun} and simplify to obtain
\begin{equation*}
\diff{A}{\theta} \sim \frac{\i\sqrt{{2\pi\rho }}B^{13/28}}{B^{1/4}} \exp\left(\frac{\rho}{\sqrt{B}}(\e^{\i\theta} - \i\theta - 1) - \i\theta(2\omega -13/14 - 1)\right).
\end{equation*}
The right-hand side of this expression is exponentially small in the limit that $B \rightarrow 0$, except on curves satisfying $\theta = 0$. These curves are the Stokes lines, across which $A$ varies rapidly. We expand locally in the neighbourhood of the Stokes lines, setting $\hat{\theta} = B^{1/4}\theta$, which gives $\mathrm{d}A/\mathrm{d}\hat{\theta}\sim \i\sqrt{{2\pi \rho }}B^{13/28}\e^{-\rho \hat{\theta}^2/2}$.
By integration, we obtain
\begin{equation}\label{0:Asmooth}
A \sim \i\sqrt{{2\pi }}B^{13/28}\int_{-\inf}^{\hat{\theta} \sqrt{\rho}}\e^{-t^2/2} \d t + A_0,
\end{equation}
where $A_0$ is a constant of integration.

In Section \ref{S:Stokes Structure}, we study the Stokes lines that follow $\theta = 0$, or $\mathrm{Im}[\chi_U] = 0$ and $\mathrm{Re}[\chi_U] > 0$. We will later establish that the physical solution requires that the remainder contribution is zero on the negative side of this Stokes line, which corresponds to choosing $A_0 = 0$. Consequently, as the Stokes line is crossed, the variation in the Stokes multiplier is found by evaluating this integral in the limit that $\hat{\theta} \rightarrow \inf$, giving
%\begin{equation}
%[A]_-^+ \sim \frac{2\pi\i}{B^{\gamma/2}}.
%\end{equation}
%We therefore define $A(\zeta) = 2\pi\i (\mathcal{S}_0 + \mathcal{S}(\zeta))/B^{\gamma/2}$, where $\mathcal{S}$ switches from zero to one as the Stokes line is crossed, and $2\pi \i \mathcal{S}_0/B^{\gamma/2} = A_0$ is the value of the Stokes multiplier on the negative side of the Stokes line. This gives the quantity switched on as the Stokes line is crossed as
%\begin{equation}
%R_N \sim \frac{2\pi\i}{B^{\gamma/2}}(\mathcal{S}_0 + \mathcal{S}(\zeta))\, F(\zeta)\,\e^{-\chi(\zeta)/\sqrt{B}},
%\end{equation}
\begin{equation*}
R_N \sim {2\pi\i }B^{13/28}\mathcal{S}(\zeta)\, F(\zeta)\,\e^{-\chi(\zeta)/\sqrt{B}},
\end{equation*}
in the limit that $B \rightarrow 0$, where $\mathcal{S}$ switches rapidly from zero to one as the Stokes line is crossed from $\theta < 0$ to $\theta > 0$. It is obtained by scaling $A$ from \eqref{0:Asmooth} to lie between zero and one.

\subsection{Summary}

If we repeat this analysis for the contribution associated with the singularity at $\zeta = -\i/q$, we can determine the equivalent late-order terms and Stokes switching behaviour associated with $\chi_L$. In this case, we find that $F(\zeta)$ is also given by \eqref{0:prefactor}, and that the Stokes switching occurs across a different Stokes line, satisfying $\mathrm{Im}[\chi_L] = 0$ with the condition that $\mathrm{Re}[\chi_L] = 0$. This curve is also studied in more detail in Section \ref{S:Stokes Structure}, and it is found that the contribution is switched as the curve is crossed from the negative side to the positive side.

The combined remainder term is given by
%\begin{equation}
%R_N \sim {2\pi\i B^{13/28}}(\mathcal{S}_{0,U} + \mathcal{S}_U(\zeta))\, F(\zeta)\,\e^{-\chi_U(\zeta)/\sqrt{B}} + \frac{2\pi\i}{B^{\gamma/2}}(\mathcal{S}_{0,L} + \mathcal{S}_L(\zeta))\, F(\zeta)\,\e^{-\chi_L(\zeta)/\sqrt{B}}.
%\end{equation}
%In this expression, the prefactor contribution is identical for both exponential contributions, but the Stokes multiplier can change. We will find in Section \ref{0:Stokes} that both $\mathcal{S}_{0,L}$ and $\mathcal{S}_{0,U}$ must be zero. Consequently, there are two exponential contributions to the late-order terms of $f(\zeta)$, associated with the essential singularities in $f(\zeta)$ at $\zeta = \pm\i/q$, which switch in behaviour rapidly across corresponding Stokes lines, with the form
\begin{equation}\label{3:remainder}
R_N \sim {2\pi\i B^{13/28} F(\zeta)}\left(\mathcal{S}_U(\zeta)\e^{-\chi_U(\zeta)/\sqrt{B}} + \mathcal{S}_L(\zeta)\e^{-\chi_L(\zeta)/\sqrt{B}}\right).
\end{equation}
Repeating this analysis for the late-order terms of $\overline{f}_n(1/\zeta)$ gives the complex conjugate expression. Consequently, the full exponentially small contribution $f_{\mathrm{exp}}$, containing all four contributions by
\begin{equation}\label{3:remainder2}
f_{\mathrm{exp}} \sim {2\pi\i B^{13/28} F(\zeta)}\left(\mathcal{S}_U(\zeta)\e^{-\chi_U(\zeta)/\sqrt{B}} + \mathcal{S}_L(\zeta)\e^{-\chi_L(\zeta)/\sqrt{B}}\right) + \overline{R}_N(1/\zeta),
\end{equation}
as $B \rightarrow 0$.

\section{Stokes structure}\label{S:Stokes Structure}

\subsection{Analysing regions of the complex plane}

We recall that active Stokes lines satisfy $\mathrm{Im}[\chi(\zeta)] = 0$ and $\mathrm{Re}[\chi(\zeta)] > 0$. Inactive Stokes lines satisfy $\mathrm{Im}[\chi(\zeta)] = 0$, but have $\mathrm{Re}[\chi(\zeta)]<0$ instead, which results in an exponentially large asymptotic contribution that does not demonstrate Stokes switching behaviour. We will also consider anti-Stokes lines, which satisfy $\mathrm{Re}[\chi(\zeta)] = 0$; as anti-Stokes lines are crossed, the associated remainder contribution switches from exponentially small behaviour to exponentially large behaviour in the asymptotic limit.

\begin{figure}[tb]
\centering
\includegraphics[width=0.8\textwidth]{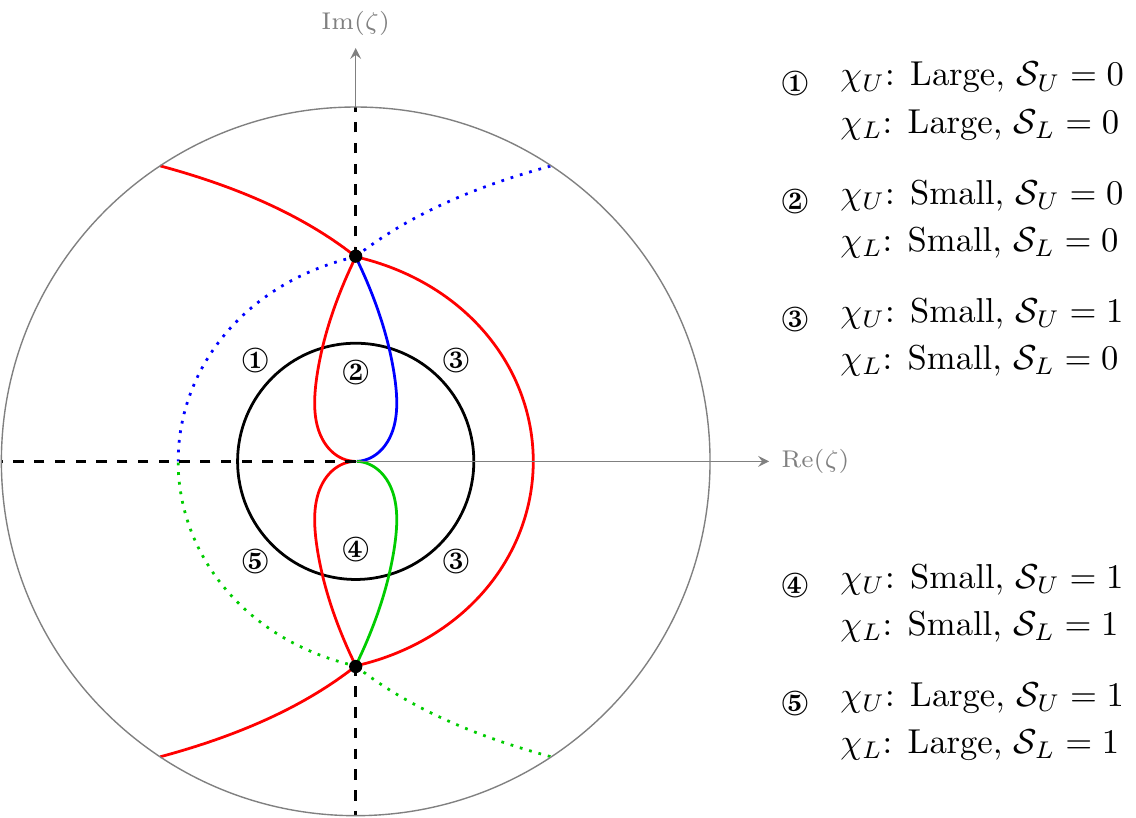}

\caption{The Stokes structure of $f(\zeta)$. Singularities in the analytically-continued bubble boundary are denoted by solid black dots. Blue and green solid curves denote Stokes lines associated with the upper and lower singularities, respectively. Anti-Stokes lines are represented as red curves and inactive Stokes lines are denoted by dotted curves. Dashed lines represent branch cuts. The physical boundary of the bubble, shown as a black circle, is divided into five regions, numbered 1--5. The exponential contributions associated with each singularity are present in regions where $\mathcal{S} = 1$, and absent in regions where $\mathcal{S}=0$. It is not possible for either $\mathcal{S}_U$ or $\mathcal{S}_L$ to be zero in both Region 1 and Region 5, where the exponential terms cause the solution to become non-physical. Consequently, solutions to the original problem can only exist if both exponential contributions cancel each other precisely.}
\label{Fig:Stokes}
\end{figure}

We find that there are two singulants, given in \eqref{0:singulants}, associated with singularities in the upper and lower half-planes, respectively. From these singulants, we apply the Stokes and anti-Stokes line criteria in order to determine the location of the corresponding Stokes lines and anti-Stokes lines. The results of this calculation are illustrated in Figure \ref{Fig:Stokes}.

In this figure, we see there are two relevant singularities, located at $\pm \i/q$. The physical boundary of the bubbles is denoted by a black circle that follows $|\zeta| = 1$. Distinct Stokes lines begin at the upper and lower singularities (shown as solid blue and green curves, respectively), while one set of anti-Stokes lines connects both of the singularities (shown as a red curve). Inactive Stokes lines (where the exponential is large, and therefore no switching occurs) are indicated by dotted blue and green lines. The exponential contribution associated with the upper and lower singularities switch in behaviour across the corresponding Stokes line, while both contributions change between being exponentially large and exponentially small as the anti-Stokes lines are crossed.

We have specified five distinct regions of the bubble surface, denoted as Regions 1 to 5. In Region 1, both exponential contributions are large. The presence of an exponentially large contribution would cause the asymptotic expansion to break down, producing non-physical results. Hence, we conclude that the Stokes multipliers must be zero in this region, leading to neither contribution being switched on. As we continue around the bubble boundary into Region 2, we cross an anti-Stokes line, and the contribution sizes change from being exponentially large to exponentially small. If we then follow the boundary into Region 3, we cross a Stokes line that switches on the exponentially small contribution associated with the upper singularity, which is therefore present in this region.

Continuing into Region 4, we cross another Stokes line, which switches on the exponentially small contribution associated with the lower singularity. Hence, both singularities are present in this region. Finally, we cross an anti-Stokes line into Region 5. This argument indicates that both contributions become exponentially large in Region 5, and the asymptotic solution breaks down. We therefore arrive at a contradiction, which appears to indicate that there cannot exist any solutions to the system at all, as any asymptotic solution must necessarily break down as an anti-Stokes line is crossed.

\subsection{Solvability condition}

Fortunately, we can resolve this apparent contradiction by observing that if the contributions associated with the upper and lower singularities exactly cancel, then the sum of the two contributions is zero, and we can continue the solution behaviour smoothly around the entire bubble boundary. Hence, we see from \eqref{3:remainder} that for solutions to exist, we require
\begin{equation*}%\label{4:solcondition}
\e^{-\chi_U(\zeta)/\sqrt{B}} + \e^{-\chi_L(\zeta)/\sqrt{B}} = 0.
\end{equation*}
That is, we find that solutions can only exist if
\begin{equation}\label{0:intcond}
\frac{\chi_U}{\sqrt{B}} - (2m+1)\pi\i= \frac{\chi_L}{\sqrt{B}},
\end{equation}
where $m$ is any positive integer. The sign of the odd integer $2m+1$ is chosen here to be negative, in order for the branches to match the labels used in \cite{GreenLustriMcCue}.

We see from \eqref{0:intcond} that, in order for a physically realistic solution to exist, $\chi_U$ and $\chi_L$ must differ by a discrete set of values that depend on the surface tension parameter, $B$.  However, by performing a simple contour integral, we have already derived the expression (\ref{4:intrelation}), which provides the difference between the singulants in terms of $U$.  Therefore, combining \eqref{4:intrelation} with \eqref{0:intcond} gives the condition we require for solutions to exist (in the small $B$ limit), namely
\begin{equation}
%\i\sqrt{U} \int_{-\i/q}^{\i/q} \frac{(z_0(s))^{3/4}(\overline{z}'_0(1/s))^{1/4}}{s^{1/2}} \d s = -\sqrt{B}(2m+1)\pi\i.
\int_{-\i/q}^{\i/q} \chi'(s) \d s = -\sqrt{B}(2m+1)\pi\i,
\label{eq:solvability}
\end{equation}
where $\chi'(\zeta)$ is given by (\ref{eqn:chidash}).  Simplifying this integral, we obtain
\begin{equation}\label{4:fullcriteria}
\i\sqrt{U} \int_{-\i/q}^{\i/q} \frac{((2-U)+Us^2)^{1/4}(U+(2-U)s^2)^{3/4}}{2s^{1/2}\sqrt{U-1}} \d s = -\sqrt{B}(2m+1)\pi\i.
\end{equation}	
This key equation provides the relationship between the surface tension $B$ and the bubble speed $U$ for each solution branch $m$. Evaluating the condition numerically shows that this condition produces solution branches for $m \geq 1$.

\begin{figure}[p]
\centering
\includegraphics[width=0.8\textwidth]{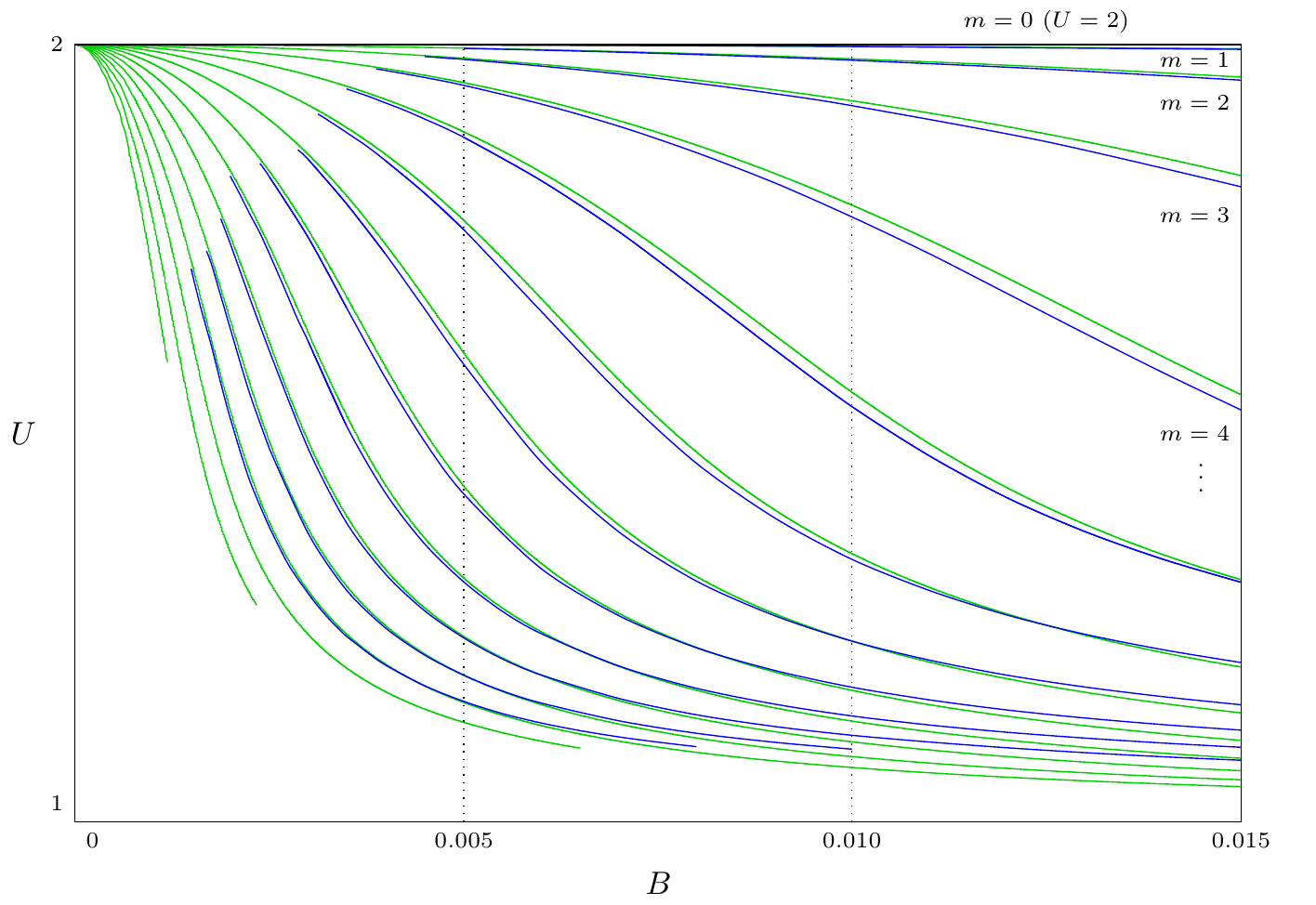}
\caption{Comparison between numerically-calculated solution branches (blue) from \cite{GreenLustriMcCue} and asymptotically-calculated solution branches from (\ref{4:fullcriteria}) (green) for solution branches $m=0,1,2,\ldots, 11$.  For each branch we see that $U\rightarrow 2^-$ as $B\rightarrow 0$.}
\label{Fig:Comparison}
\end{figure}

We compare the  solvability condition (\ref{4:fullcriteria}) with our numerical results from \cite{GreenLustriMcCue} in Figure~\ref{Fig:Comparison} for the first dozen solution branches.  The comparison is remarkably good, showing that the asymptotic result provides accurate approximations to the solution branches, especially in the limit that $B \rightarrow 0$ (and hence $U \rightarrow 2^-$).  Note the numerical scheme in \cite{GreenLustriMcCue} was not able to compute solutions for very small values of $B$; however, the asymptotic result (\ref{4:fullcriteria}) works, of course, for all $B>0$.

\subsection{Simple scaling law}

To determine the asymptotic behaviour of the branches in Figure \ref{Fig:Comparison} in the limit $B\rightarrow 0$, we perform an asymptotic analysis of the integral \eqref{4:fullcriteria} in the limit that $U \rightarrow 2^-$.   In this limit, we see that $1/q \sim \i \sqrt{2/(2-U)}$. This scaling suggests defining a new quantity $s' = \sqrt{2-U}\, s$, and studying the integral expressed in terms of $s'$. This will accurately capture the contributions to the integral in \eqref{4:fullcriteria} from the end points, which dominate the behaviour in the limit $U \rightarrow 2^-$.  The condition \eqref{4:fullcriteria} becomes, in this limit,
\begin{equation}
\frac{\i (2-U)^{1/4}}{2^{1/4}} \int_{-\i\sqrt{2}}^{\i\sqrt{2}} \left[\frac{(-1-2/s')^{3/4}(-s'^2)^{1/4}}{\sqrt{s'}} \right]\d s' \sim -\sqrt{B}(2m+1)\pi\i,
\label{eq:integral}
\end{equation}
where omitted terms on the left-hand side of \eqref{eq:integral} are $\mathcal{O}((2-U)^{5/4})$.

Evaluating this integral \eqref{eq:integral} to leading order using hypergeometric functions gives the condition
\begin{equation*}%\label{4:scaling}
\frac{2^{13/4}\Gamma(7/4)}{\sqrt{\pi}\Gamma(1/4)}(2-U)^{1/4} \sim (2m+1)\sqrt{B} \quad \mathrm{as} \quad U \rightarrow 2^-,
\end{equation*}
which is equivalent to (\ref{eq:mainresult}).   This scaling law for the solution branches in the small-surface-tension limit confirms the predictions from \cite{GreenLustriMcCue,HongFamily88} that $U\sim 2- k B^{2}$, but goes further by calculating the value of $k$ explicitly in terms of $m$.  %As a check on how close (\ref{4:scaling}) is to the full solvability condition (\ref{4:fullcriteria}), we plot both in Figure \ref{Fig:ScalingLaw} for small values of $B$. In this figure we see that the scaling law increases in accuracy near $U = 2$, correctly describing the asymptotic behaviour of the solution branches in this asymptotic limit, as expected.

%\begin{figure}[p]
%\centering
%\caption{Comparison of the solution branches obtained using the asymptotic condition \eqref{4:fullcriteria}, depicted in green, against the scaling law \eqref{4:scaling}, depicted in red. The accuracy of the scaling law visibly improves on each branch as $U \rightarrow 2^-$, which is a consequence of the fact that the scaling law was derived in this asymptotic limit.}
%\label{Fig:ScalingLaw}
%\end{figure}

\section{Solution in the physical plane}\label{sec:bubbleshape}

\subsection{Effect of remainder terms on bubble shape}

Equation \eqref{3:remainder2} provides an asymptotic expression for the remainder term $f_{\mathrm{exp}}$ for our mapping function $z(\zeta)=z_0(\zeta)+f(\zeta)$, valid after truncating the expansion (\ref{0:fseries}) optimally.  While the remainder term in \eqref{3:remainder2} is exponentially small in the limit $B\rightarrow 0$, and therefore formally smaller than each of the algebraic terms in (\ref{0:fseries}), we argue that it is this term that is responsible for the non-convex exotic bubble shapes demonstrated numerically in \cite{GreenLustriMcCue}.  Indeed, as $B$ increases, the wavelike nature of $f_{\mathrm{exp}}$ plays an important role in the bubble shape, as we shall now explain.

We will define an approximate expression for the map $z(\zeta)$ by
\begin{equation}\label{5:zapp}
z(\zeta) \approx z_{\mathrm{app}}(\zeta) = z_0(\zeta) + f_{\mathrm{exp}}(\zeta).
\end{equation}
This approximation involves the zero-surface-tension map $z_0$ plus the exponentially small remainder term $f_{\mathrm{exp}}$, but neglects all of the algebraic terms in $Bf_1$, $B^2f_2$, and so on.  As such, it is important to emphasize that this approximation cannot be used to make reliable quantitative predictions, as we are omitting algebraic terms that are larger than $f_{\mathrm{exp}}$ for $B\ll 1$.

%We note that $f_{\mathrm{exp}}$ varies smoothly in the manner described in \eqref{0:Asmooth} with $A_0 = 0$.

\begin{figure}
\centering
\subfloat[$m=1$]{
\includegraphics{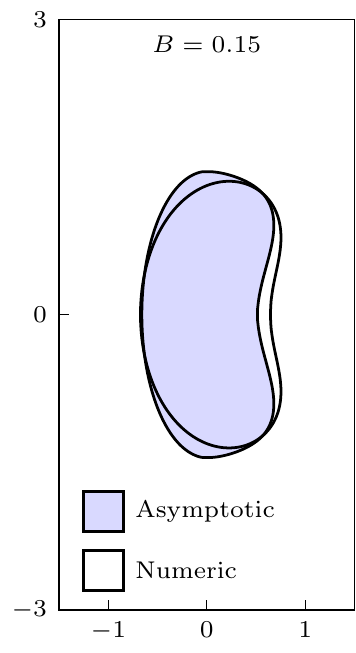}
%\begin{tikzpicture}
%[xscale=1,>=stealth,yscale=1]
%
%\fill[blue,opacity=0.15] plot[smooth] file {m1_B0p15_U1p5.txt};
%\draw[black,thick] plot[smooth] file {m1_B0p15_U1p5.txt};
%\draw[black,thick] plot[smooth] file {m1_comp.txt};
%
%\node at (0,2.75) {\scriptsize{$B = 0.15$}};
%
%\node at (-1.5,3) [left] {\scriptsize{$3$}};
%\draw (-1.5,0) node[left] {\scriptsize{$0$}} -- (-1.4,0);
%\node at (-1.5,-3) [left] {\scriptsize{$-3$}};
%\node at (-1,-3) [below] {\scriptsize{$-1$}};
%\draw (0,-3) node[below] {\scriptsize{$0$}} -- (0,-2.9) ;
%\node at (1,-3) [below] {\scriptsize{$1$}};
%\draw (-1,-3) -- (-1,-2.9);
%\draw (1,-3) -- (1,-2.9);
%
%\draw (-1.5,3)  -- (1.5,3) -- (1.5,-3) -- (-1.5,-3) -- (-1.5,0) -- cycle;
%
%\fill[blue,opacity=0.15]  (-1.25,-1.8) -- (-0.85,-1.8) -- (-0.85,-2.2) -- (-1.25,-2.2) -- cycle;
%\draw[thick]  (-1.25,-1.8) -- (-0.85,-1.8) -- (-0.85,-2.2) -- (-1.25,-2.2) -- cycle;
%
%\draw[thick]  (-1.25,-2.4) -- (-0.85,-2.4) -- (-0.85,-2.8) -- (-1.25,-2.8) -- cycle;
%\node at (-0.85,-2) [right] {\scriptsize{Asymptotic}};
%\node at (-0.85,-2.6) [right] {\scriptsize{Numeric}};

%\end{tikzpicture}
}
\subfloat[$m=2$]{
\includegraphics{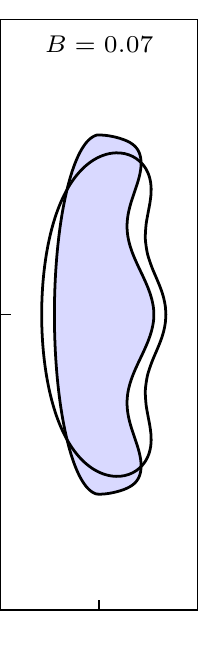}
%\begin{tikzpicture}
%[xscale=1,>=stealth,yscale=1]
%
%\fill[blue,opacity=0.15] plot[smooth] file {m2_B0p07_U1p35.txt};
%
%\node[white] at (0,-3) [below] {\scriptsize{$-1$}};
%\node[white] at (0,3) {\scriptsize{$3$}};
%\draw (-1,0) -- (-0.9,0);
%\draw (0,-3) -- (0,-2.9) ;
%
%\draw[black,thick] plot[smooth] file {m2_B0p07_U1p35.txt};
%\draw[black,thick] plot[smooth] file {m2_comp.txt};
%\node at (0,2.75) {\scriptsize{$B = 0.07$}};
%\draw (-1.0,3) -- (1.0,3) -- (1.0,-3) -- (-1.0,-3) -- cycle;
%\end{tikzpicture}
}
\subfloat[$m=3$]{
\includegraphics{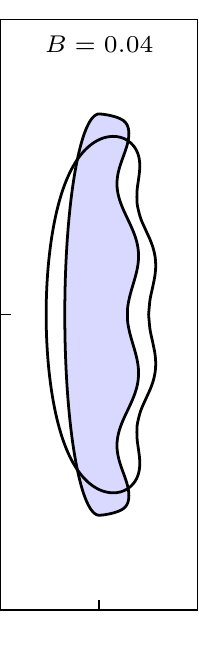}
%\begin{tikzpicture}
%[xscale=1,>=stealth,yscale=1]
%
%\fill[blue,opacity=0.15] plot[smooth] file {m3_B0p04_U1p3.txt};
%\draw[black,thick] plot[smooth] file {m3_B0p04_U1p3.txt};
%\draw[black,thick] plot[smooth] file {m3_comp.txt};
%
%\node[white] at (0,-3) [below] {\scriptsize{$-1$}};
%\node[white] at (0,3) {\scriptsize{$3$}};
%\draw (-1,0) -- (-0.9,0);
%\draw (0,-3) -- (0,-2.9) ;
%
%\node at (0,2.75) {\scriptsize{$B = 0.04$}};
%\draw (-1.0,3) -- (1.0,3) -- (1.0,-3) -- (-1.0,-3) -- cycle;
%\end{tikzpicture}
}
\subfloat[$m=4$]{
\includegraphics{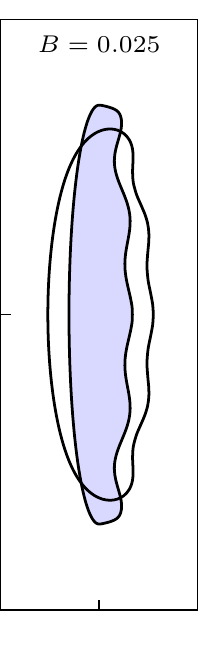}
%\begin{tikzpicture}
%[xscale=1,>=stealth,yscale=1]
%
%\node[white] at (0,-3) [below] {\scriptsize{$-1$}};
%\node[white] at (0,3) {\scriptsize{$3$}};
%\draw (-1,0) -- (-0.9,0);
%\draw (0,-3) -- (0,-2.9) ;
%
%\fill[blue,opacity=0.15] plot[smooth] file {m4_B0p025_U1p3.txt};
%\draw[black,thick] plot[smooth] file {m4_B0p025_U1p3.txt};
%\draw[black,thick] plot[smooth] file {m4_comp.txt};
%
%\node at (0,2.75) {\scriptsize{$B = 0.025$}};
%\draw (-1.0,3) -- (1.0,3) -- (1.0,-3) -- (-1.0,-3) -- cycle;
%\end{tikzpicture}
}
\subfloat[$m=7$]{
\includegraphics{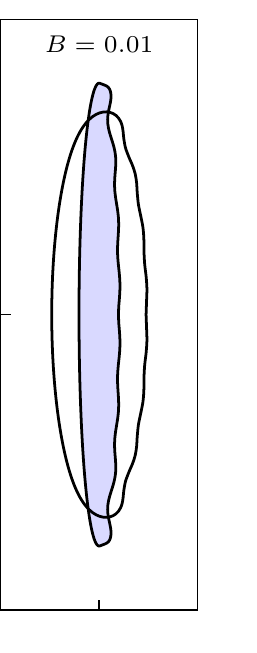}
%\begin{tikzpicture}
%[xscale=1,>=stealth,yscale=1]
%
%\node[white] at (0,-3) [below] {\scriptsize{$-1$}};
%\node[white] at (0,3) {\scriptsize{$3$}};
%\node[white] at (1,-3) [right] {\scriptsize{$-3$}};
%\draw (-1,0) -- (-0.9,0);
%\draw (0,-3) -- (0,-2.9) ;
%
%\fill[blue,opacity=0.15] plot[smooth] file {m7_B0p01_U1p2.txt};
%\draw[black,thick] plot[smooth] file {m7_B0p01_U1p2.txt};
%\draw[black,thick] plot[smooth] file {m7_comp.txt};
%
%\node at (0,2.75) {\scriptsize{$B = 0.01$}};
%\draw (-1.0,3) -- (1.0,3) -- (1.0,-3) -- (-1.0,-3) -- cycle;
%\end{tikzpicture}
}
\caption{Approximate bubble solutions drawn using \eqref{5:zapp}, depicted as black curves containing shaded regions, for a range of solution branches. We see that the bubbles display qualitative agreement with the numerical predictions from \cite{GreenLustriMcCue}, depicted as black curves without internal shading. In particular, the exponentially small contribution causes a series of ripples with $m$ non-convex peaks on the bubble surface.  Values of $B$ were selected so that the ripples are visible and may be compared directly. 
%Each plot is illustrated over an identical region in the physical plane.
}
\label{Fig:Bubble1}
\end{figure}

Approximate bubble solutions computed using \eqref{5:zapp} are depicted in Figure \ref{Fig:Bubble1} for a range of different surface tension values and solution branch numbers $m$.  It is apparent that the choice of $m$ determines the number of oscillations that are present in the region between the Stokes lines, which translates to the number of convex ripples on the surface of the bubble. As $m$ increases, the bubble becomes narrower, and contains more ripples on the surface. These results show strong qualitative agreement with the numerical results from \cite{GreenLustriMcCue}.

%While the bubbles do appear to be consistent quantitatively, it is important to recall that the approximation \eqref{5:zapp} omits several terms that would be expected to play visible roles in the solution. We therefore focus on the qualitative agreement, particularly noting the consistent number of ripples, the vertical symmetry of the bubbles, and the direction of the tips on the bubble surface.

\begin{figure}
\centering
\subfloat{
\includegraphics{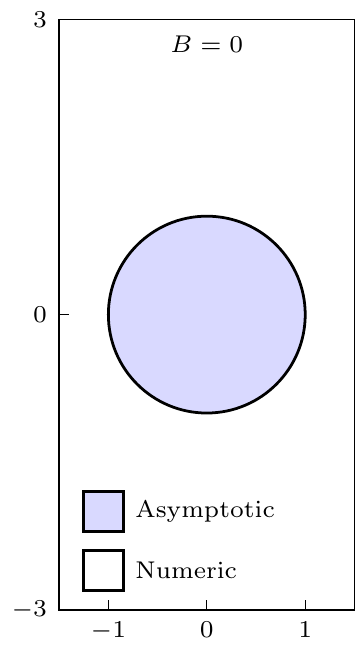}
%\begin{tikzpicture}
%[xscale=1,>=stealth,yscale=1]
%
%\fill[blue,opacity=0.15] (0,0) circle (1);
%\draw[black,thick] (0,0) circle (1);
%
%\node at (0,2.75) {\scriptsize{$B = 0$}};
%
%\node at (-1.5,3) [left] {\scriptsize{$3$}};
%\draw (-1.5,0) node[left] {\scriptsize{$0$}} -- (-1.4,0);
%\node at (-1.5,-3) [left] {\scriptsize{$-3$}};
%\node at (-1,-3) [below] {\scriptsize{$-1$}};
%\draw (0,-3) node[below] {\scriptsize{$0$}} -- (0,-2.9) ;
%\node at (1,-3) [below] {\scriptsize{$1$}};
%\draw (-1,-3) -- (-1,-2.9);
%\draw (1,-3) -- (1,-2.9);
%
%\draw (-1.5,3)  -- (1.5,3) -- (1.5,-3) -- (-1.5,-3) -- (-1.5,0) -- cycle;
%
%\fill[blue,opacity=0.15]  (-1.25,-1.8) -- (-0.85,-1.8) -- (-0.85,-2.2) -- (-1.25,-2.2) -- cycle;
%\draw[thick]  (-1.25,-1.8) -- (-0.85,-1.8) -- (-0.85,-2.2) -- (-1.25,-2.2) -- cycle;
%
%\draw[thick]  (-1.25,-2.4) -- (-0.85,-2.4) -- (-0.85,-2.8) -- (-1.25,-2.8) -- cycle;
%\node at (-0.85,-2) [right] {\scriptsize{Asymptotic}};
%\node at (-0.85,-2.6) [right] {\scriptsize{Numeric}};
%\end{tikzpicture}
}
\subfloat{
\includegraphics{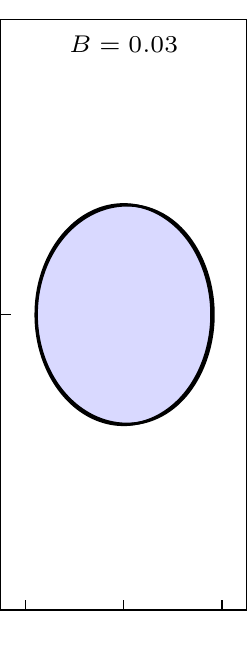}
%\begin{tikzpicture}
%[xscale=1,>=stealth,yscale=1]
%
%\fill[blue,opacity=0.15] plot[smooth] file {m2_B0p03_U1p8.txt};
%\draw[black,thick] plot[smooth] file {m2_B0p03_U1p8.txt};
%\draw[black,thick] plot[smooth] file {m23_comp.txt};
%
%\node at (0,2.75) {\scriptsize{$B = 0.03$}};
%
%\node[white] at (0,-3) [below] {\scriptsize{$-1$}};
%\node[white] at (0,3) {\scriptsize{$3$}};
%\draw (-1.25,0) -- (-1.15,0);
%\draw (0,-3) -- (0,-2.9) ;
%\draw (1,-3)  -- (1,-2.9) ;
%\draw (-1,-3)  -- (-1,-2.9) ;
%
%\draw (-1.25,3)  -- (1.25,3) -- (1.25,-3) -- (-1.25,-3) -- (-1.25,0) -- cycle;
%\end{tikzpicture}
}
\subfloat{
\includegraphics{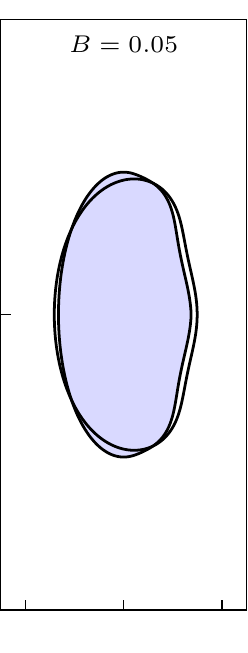}
%\begin{tikzpicture}
%[xscale=1,>=stealth,yscale=1]
%
%\fill[blue,opacity=0.15] plot[smooth] file {m2_B0p05_U1p5.txt};
%\draw[black,thick] plot[smooth] file {m25_comp.txt};
%
%\node[white] at (0,-3) [below] {\scriptsize{$-1$}};
%\node[white] at (0,3) {\scriptsize{$3$}};
%\draw (-1.25,0) -- (-1.15,0);
%\draw (0,-3) -- (0,-2.9) ;
%\draw (1,-3)  -- (1,-2.9) ;
%\draw (-1,-3)  -- (-1,-2.9) ;
%
%\draw[black,thick] plot[smooth] file {m2_B0p05_U1p5.txt};
%\node at (0,2.75) {\scriptsize{$B = 0.05$}};
%\draw (-1.25,3)  -- (1.25,3) -- (1.25,-3) -- (-1.25,-3) -- (-1.25,0) -- cycle;
%\end{tikzpicture}
}
\subfloat{
\includegraphics{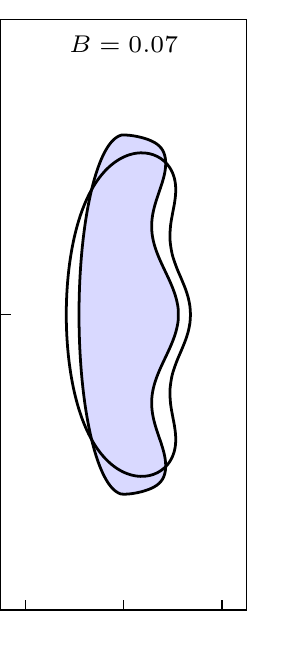}
%\begin{tikzpicture}
%[xscale=1,>=stealth,yscale=1]
%\draw[white] (1.25,0) node[right] {\scriptsize{$0$}} -- (1.15,0);
%\fill[blue,opacity=0.15] plot[smooth] file {m2_B0p07_U1p35.txt};
%\draw[black,thick] plot[smooth] file {m2_B0p07_U1p35.txt};
%\draw[black,thick] plot[smooth] file {m2_comp.txt};
%
%\node[white] at (0,-3) [below] {\scriptsize{$-1$}};
%\node[white] at (0,3) {\scriptsize{$3$}};
%\draw (-1.25,0) -- (-1.15,0);
%\draw (0,-3) -- (0,-2.9) ;
%\draw (1,-3)  -- (1,-2.9) ;
%\draw (-1,-3)  -- (-1,-2.9) ;
%
%\node at (0,2.75) {\scriptsize{$B = 0.07$}};
%\draw (-1.25,3)  -- (1.25,3) -- (1.25,-3) -- (-1.25,-3) -- (-1.25,0) -- cycle;
%\end{tikzpicture}
}
\caption{Approximate bubble solutions from the branch $m=2$ drawn using \eqref{5:zapp}, depicted as black curves containing shaded regions, for a range of different $B$ values. As the surface tension increases in size, the ripples on the surface of the bubble also increase. The asymptotic predictions are compared with numerical solutions from \cite{GreenLustriMcCue}, depicted as black curves without internal shading.}% The qualitative agreement with the overall bubble shape increases for smaller values of $B$, as expected.} % While we see qualitative agreement in the ripple behaviour with numerical solutions \cite{GreenLustriMcCue} for the smaller values of $B$, the approximation for $B = 0.09$ displays large ripples and edges with more curvature than predicted by the numerical results. This suggests that $B$ is too large in this case for the asymptotic approximation to provide a useful quantitative description of the surface behaviour.}
\label{Fig:Bubble2}
\end{figure}

In Figure \ref{Fig:Bubble2}, we illustrate the effect of following a particular branch of the solution; in the case, we select $m = 2$. We see that as the surface tension decreases, the ripples become smaller and less visible. This is expected, as the amplitude of the oscillations is exponentially small as $B \rightarrow 0$. Again, these results are consistent with the numerical predictions from \cite{GreenLustriMcCue}.

%For larger depicted values of $B$ in Figure \ref{Fig:Bubble2}, we see that the ripples on the bubble surface are quite large, indicating that the small $B$ approximation is unlikely to be valid, and therefore this does not produce a quantitatively accurate approximation of the bubble surface. This can be confirmed by comparing the asymptotic results for $B = 0.09$ with the numerical results in \cite{GreenLustriMcCue}.

\subsection{Number of tips appearing on nonconvex bubbles}

We now perform an analysis to explain the number of tips that appear on bubble shapes along solution branches.  We denote the point at which the Stokes line from $\zeta=\i/q$ intersects the unit circle by $\zeta=\zeta_{S^+}$.  We make the important observation that the wavelike behaviour is caused by the form of the remainder term in Region 3 (see figure~\ref{Fig:Stokes}), which includes the oscillatory factor  $\e^{-\i\,\mathrm{Im}[\chi_U]/\sqrt{B}}$.  Here, $\mathrm{Im}[\chi_U]$ is a monotonically decreasing function along this part of the unit circle from $\zeta=1$ to $\zeta=\zeta_{S^+}$, at which point it must vanish (since Stokes lines require that $\mathrm{Im}[\chi]=0$).  Therefore, for the top half of the bubble, we can count the number of wavelengths of this oscillatory factor as being
\begin{equation*}
\frac{1}{2\pi}
\left(
\frac{\mathrm{Im}[\chi_U(1)]}{\sqrt{B}}-\frac{\mathrm{Im}[\chi_U(\zeta_{S^+})]}{\sqrt{B}}
\right),
\end{equation*}
and the number of wavelengths along the entire Region 3 as twice this value.  However, from \eqref{eq:chiU1chiL1} and \eqref{eq:solvability}, we see that
\begin{equation*}
\chi_U(1)=2\sqrt{B}(2m+1)\pi\i.
%\label{eq:chi1}
\end{equation*}
Therefore, we determine that the number of wavelengths of the oscillatory contribution as being
\begin{equation*}
\frac{1}{\pi\sqrt{B}}\mathrm{Im}[\chi_U(1)]=m+\frac{1}{2}.
%\label{eq:numberwaves}
\end{equation*}
%For $m=0$ the solution is circular so this interpretation is that there is half of a wavelength.  For $m=1$, \ldots ....  As such, our analysis explains why bubbles shapes corresponding to solutions on the $m$th branch develop $m+1$ tips.

\begin{figure}
\centering
\subfloat[Bubble from branch $m=3$]{
\includegraphics{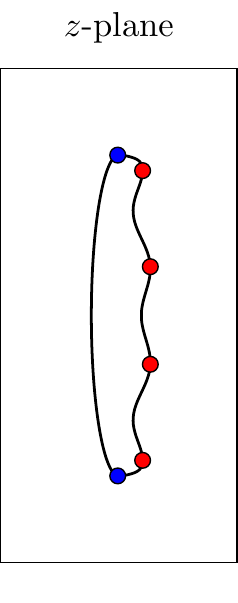}
%\begin{tikzpicture}
%[xscale=0.8*1,>=stealth,yscale=0.8*1]
%\draw[white,->] (0,3.14) -- (0,3.4) node [above] {{$\theta$}};
%\node[white] at (0.015,-3.14) [below] {\scriptsize{$0.015$}};
%
%\node at (0,3.3) [above] {{$z$-plane}};
%
%\draw[black,thick] plot[smooth] file {m3_B0p04_U1p3.txt};
%
%\fill[red] (0.2882,0) circle (0.1);
%\draw (0.2882,0) circle (0.1);
%\fill[red] (0.1817,1.34) circle (0.1);
%\draw (0.1817,1.34) circle (0.1);
%\fill[red] (0.1817,-1.34) circle (0.1);
%\draw (0.1817,-1.34) circle (0.1);
%\fill[blue] (-0.0138,2.0371) circle (0.1);
%\draw (-0.0138,2.0371) circle (0.1);
%\fill[blue] (-0.0138,-2.0371) circle (0.1);
%\draw (-0.0138,-2.0371) circle (0.1);
%
%%\node at (0,2.75) {\scriptsize{$B = 0.04$}};
%\draw (-1.5,3.14) -- (1.5,3.14) -- (1.5,-3.14) -- (-1.5,-3.14) -- cycle;
%\end{tikzpicture}
}
\subfloat[Mapped Plane]{
\includegraphics{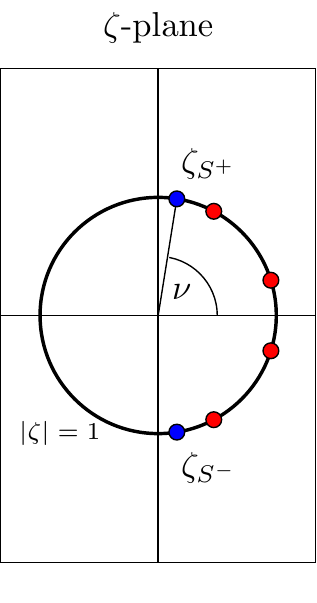}
%\begin{tikzpicture}
%[xscale=0.8*1,>=stealth,yscale=0.8*1]
%
%\draw[white,->] (0,3.14) -- (0,3.4) node [above] {{$\theta$}};
%\node[white] at (0.015,-3.14) [below] {\scriptsize{$0.015$}};
%
%\node at (0,3.3) [above] {{$\zeta$-plane}};
%
%
%\draw[line width=0.35mm] (0,0) circle (1.5);
%
%\draw(-2,0) -- (2,0);
%\draw(0,-3.14) -- (0,3.14);
%
%\draw (0,0) -- (0.2370,1.4812);
%\begin{scope}
%\clip (0,0) -- (2,0) -- (0.2811,1.4734) -- (0,0);
%\draw (0,0) circle (0.75);
%\end{scope}
%\node at (0.3,0.3) {{$\nu$}};
%\node at (-1.25,-1.5) {\scriptsize{$|\zeta| = 1$}};
%
%\fill[red] (1.4316,0.4478) circle (0.1);
%\draw (1.4316,0.4478) circle (0.1);
%\fill[red] (1.4316,-0.4478) circle (0.1);
%\draw (1.4316,-0.4478) circle (0.1);
%\fill[red] (0.7061,1.3234) circle (0.1);
%\draw (0.7061,1.3234) circle (0.1);
%\fill[red] (0.7061,-1.3234) circle (0.1);
%\draw ( 0.7061,-1.3234) circle (0.1);
%\fill[blue] (0.2370,1.4812) circle (0.1);
%\draw (0.2370,1.4812) circle (0.1);
%\fill[blue] (0.2370,-1.4812) circle (0.1);
%\draw ( 0.2370,-1.4812) circle (0.1);
%
%
%\node at (0.1370,1.5812) [above right] {{$\zeta_{S^+}$}};
%\node at (0.1370,-1.5812) [below right] {{$\zeta_{S^-}$}};
%
%\draw(-2,3.14) -- (2,3.14) -- (2,-3.14) -- (-2,-3.14) -- cycle;
%
%
%\end{tikzpicture}
}
\subfloat[Angle of Stokes intersection points and oscillation peaks in the mapped plane]{
\includegraphics{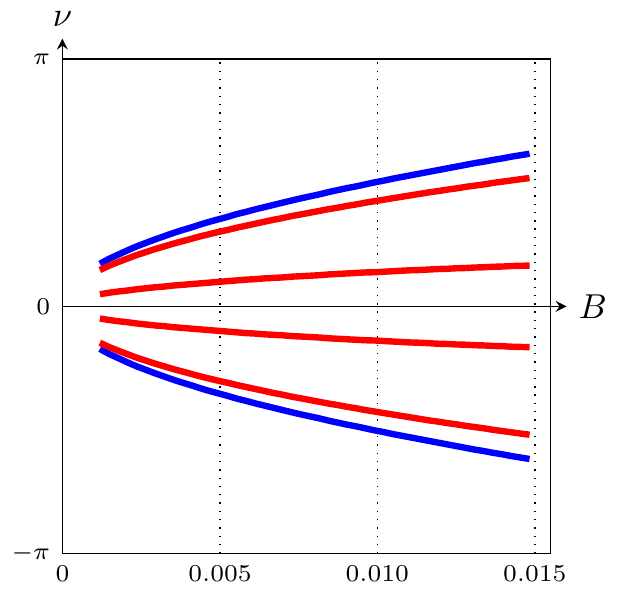}
%\begin{tikzpicture}
%[xscale=0.8*400,>=stealth,yscale=0.8*2]
%
%\node at (0,1.57) [left] {\scriptsize{$\pi$}};
%\node at (0,-1.57) [left] {\scriptsize{$-\pi$}};
%\node at (0,0) [left] {\scriptsize{$0$}};
%\draw[->] (0,1.57) -- (0,1.7) node [above] {{$\nu$}};
%\draw (0,0) -- (0.0155,0);
%\draw[->] (0.0155,0) -- (0.016,0) node[right] {{$B$}};
%\node at (0,-1.57) [below] {\scriptsize{$0$}};
%\node at (0.005,-1.57) [below] {\scriptsize{$0.005$}};
%\node at (0.01,-1.57) [below] {\scriptsize{$0.010$}};
%\node at (0.015,-1.57) [below] {\scriptsize{$0.015$}};
%
%\draw [dotted] (0.005,-1.57) -- (0.005,1.57);
%\draw [dotted] (0.01,-1.57) -- (0.01,1.57);
%\draw [dotted] (0.015,-1.57) -- (0.015,1.57);
%
%\draw[blue,line width=0.65mm] plot[smooth] file {SL_tip0_3b.txt};
%\draw[blue,line width=0.65mm] plot[smooth] file {SL_tip1_3b.txt};
%
%\draw[red,line width=0.65mm] plot[smooth] file {SL_tip2_3b.txt};
%\draw[red,line width=0.65mm] plot[smooth] file {SL_tip3_3b.txt};
%\draw[red,line width=0.65mm] plot[smooth] file {SL_tip4_3b.txt};
%\draw[red,line width=0.65mm] plot[smooth] file {SL_tip5_3b.txt};
%
%\draw(0,1.57) -- (0.0155,1.57) -- (0.0155,-1.57) -- (0,-1.57) -- cycle;
%
%\end{tikzpicture}
}

\subfloat[Bubble from branch $m=4$]{
\includegraphics{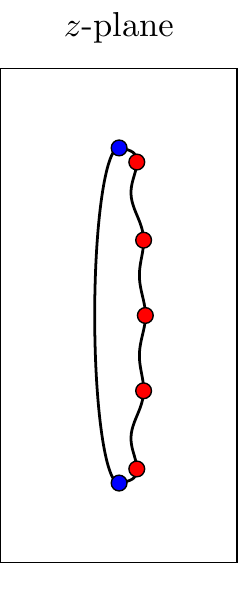}
%\begin{tikzpicture}
%[xscale=0.8*1,>=stealth,yscale=0.8*1]
%
%\draw[white,->] (0,3.14) -- (0,3.4) node [above] {{$\theta$}};
%\node[white] at (0.015,-3.14) [below] {\scriptsize{$0.015$}};
%
%\node at (0,3.3) [above] {{$z$-plane}};
%
%\draw[black,thick] plot[smooth] file {m4_B0p025_U1p3.txt};
%
%\fill[red] (0.2617,0.5185) circle (0.1);
%\draw (0.2617,0.5185) circle (0.1);
%\fill[red] (0.2617,-0.5185) circle (0.1);
%\draw (0.2617,-0.5185) circle (0.1);
%\fill[red] (0.1550,1.5440) circle (0.1);
%\draw (0.1550,1.5440) circle (0.1);
%\fill[red] (0.1550,-1.5440) circle (0.1);
%\draw (0.1550,-1.5440) circle (0.1);
%\fill[blue] (0.0040,2.1280) circle (0.1);
%\draw (0.0040,2.1280) circle (0.1);
%\fill[blue] (0.0040,-2.1280) circle (0.1);
%\draw (0.0040,-2.1280) circle (0.1);
%
%%\node at (0,2.75) {\scriptsize{$B = 0.025$}};
%\draw (-1.5,3.14) -- (1.5,3.14) -- (1.5,-3.14) -- (-1.5,-3.14) -- cycle;
%\end{tikzpicture}
}
\subfloat[Mapped plane]{
\includegraphics{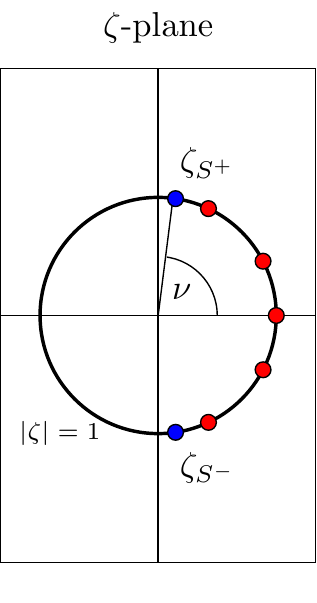}
%\begin{tikzpicture}
%[xscale=0.8*1,>=stealth,yscale=0.8*1]
%
%
%\draw[white,->] (0,3.14) -- (0,3.4) node [above] {{$\theta$}};
%\node[white] at (0.015,-3.14) [below] {\scriptsize{$0.015$}};
%
%\node at (0,3.3) [above] {{$\zeta$-plane}};
%\draw[line width=0.35mm] (0,0) circle (1.5);
%
%\draw(-2,0) -- (2,0);
%\draw(0,-3.14) -- (0,3.14);
%
%\draw (0,0) -- (0.1880,1.4882);
%\begin{scope}
%\clip (0,0) -- (2,0) -- (0.2207,1.4837) -- (0,0);
%\draw (0,0) circle (0.75);
%\end{scope}
%\node at (0.3,0.3) {{$\nu$}};
%\node at (-1.25,-1.5) {\scriptsize{$|\zeta| = 1$}};
%
%\fill[blue] (0.2207,1.4837) circle (0.1);
%\draw (0.2207,1.4837) circle (0.1);
%\fill[blue] (0.2207,-1.4837) circle (0.1);
%\draw (0.2207,-1.4837) circle (0.1);
%\fill[red] (1.5,0) circle (0.1);
%\draw (1.5,0) circle (0.1);
%\fill[red] (1.3322,0.6894) circle (0.1);
%\draw (1.3322,0.6894) circle (0.1);
%\fill[red] (1.3322,-0.6894) circle (0.1);
%\draw (1.3322,-0.6894) circle (0.1);
%\fill[red] (0.6387,1.3572) circle (0.1);
%\draw (0.6387,1.3572) circle (0.1);
%\fill[red] (0.6387,-1.3572) circle (0.1);
%\draw (0.6387,-1.3572) circle (0.1);
%
%\node at (0.1207,1.5837) [above right] {$\zeta_{S^+}$};
%\node at (0.1207,-1.5837) [below right] {$\zeta_{S^-}$};
%
%
%
%\draw(-2,3.14) -- (2,3.14) -- (2,-3.14) -- (-2,-3.14) -- cycle;
%
%
%\end{tikzpicture}
}
\subfloat[Angle of Stokes intersection points and oscillation peaks in the mapped plane]{
\includegraphics{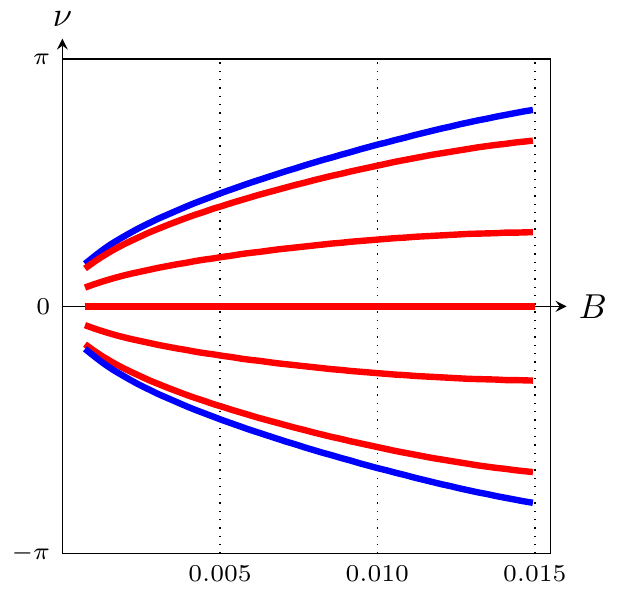}
%\begin{tikzpicture}
%[xscale=0.8*400,>=stealth,yscale=0.8*2]
%
%\node at (0,1.57) [left] {\scriptsize{$\pi$}};
%\node at (0,-1.57) [left] {\scriptsize{$-\pi$}};
%\node at (0,0) [left] {\scriptsize{$0$}};
%\draw[->] (0,1.57) -- (0,1.7) node [above] {{$\nu$}};
%\draw (0,0) -- (0.0155,0);
%\draw[->] (0.0155,0) -- (0.016,0) node[right] {{$B$}};
%\node at (0.005,-1.57) [below] {\scriptsize{$0.005$}};
%\node at (0.01,-1.57) [below] {\scriptsize{$0.010$}};
%\node at (0.015,-1.57) [below] {\scriptsize{$0.015$}};
%\draw [dotted] (0.005,-1.57) -- (0.005,1.57);
%\draw [dotted] (0.01,-1.57) -- (0.01,1.57);
%\draw [dotted] (0.015,-1.57) -- (0.015,1.57);
%
%\draw[blue,line width=0.65mm] plot[smooth] file {SL_tip0_4b.txt};
%\draw[red,line width=0.65mm] plot[smooth] file {SL_tip1_4b.txt};
%\draw[red,line width=0.65mm] plot[smooth] file {SL_tip2_4b.txt};
%\draw[red,line width=0.65mm] plot[smooth] file {SL_tip3_4b.txt};
%\draw[red,line width=0.65mm] plot[smooth] file {SL_tip5_4b.txt};
%\draw[blue,line width=0.65mm] plot[smooth] file {SL_tip4_4b.txt};
%\draw[red,line width=0.65mm] (0.0007,0) -- (0.015,0);
%
%\draw(0,1.57) -- (0.0155,1.57) -- (0.0155,-1.57) -- (0,-1.57) -- cycle;
%
%\end{tikzpicture}

}

\caption{Position of tips (red circles) and Stokes intersection points (blue circles) in the physical plane and mapped plane are shown in (a)--(b) for the $m=3$ branch, and (d)--(e) for the $m=4$ branch. The location of the tips (red curves) and intersection points (blue curves) in the mapped plane as $B$ varies is depicted in (c) for the $m=3$ branch, and (f) for the $m=4$ branch. These branches are representative of the tip structure for odd and even branches respectively. In each case, the region between the Stokes intersection points contains $m+1/2$ oscillation wavelengths, and the active region narrows as $B$ varies, ensuring that this number of wavelengths remains constant along the branch.}
\label{PeakFig}
\end{figure}

In order to illustrate this behaviour, we illustrate a representative bubble for $m=3$ and $m=4$ in Figure \ref{PeakFig} (a) and (d). The corresponding locations in the mapped plane are shown in Figure \ref{PeakFig} (b) and (e). 

On the bubble corresponding with branch $m$, there are $m+1$ tips (denoted by red circles) contained in the region between the Stokes intersection points (denoted by blue circles). The tips are each separated by one wavelength, giving $m$ wavelengths between tips. There is a further quarter of a wavelength between between the uppermost tip and the Stokes intersection point $\zeta_{S^+}$ where the oscillations are switched on, and between the lowermost tip and the Stokes intersection point $\zeta_{S^-}$, giving a total of $m+1$ tips in a region containing $m + 1/2$ wavelengths. The bubble is vertically symmetric, hence if $m$ is even there is a tip on the horizontal axis, while if $m$ is odd the tips are located off the horizontal axis.

Figure \ref{PeakFig} (c) and (f) show how the angular position of the tips and the Stokes intersection points in the mapped plane as the surface tension decreases for the $m=3$ and $m=4$ solution branches. We conclude from the form of $f_{\mathrm{exp}}$ in \eqref{3:remainder2} that the wavelength must decrease as $B \rightarrow 0$. These figures confirm that the region in which the contribution is present also decreases in size in this limit. This ensures that the total number of wavelengths in the active region, and therefore also the number of tips that form on the bubble, remains constant along each branch. 

\section{Discussion}\label{sec:discussion}

We have been concerned with a model for steadily moving bubbles in an unbounded Hele-Shaw cell.  For nonzero surface tension values $B>0$, this model gives rise to a countable set of nonunique solutions that we have labelled $m=1,2,\dots$,  in addition to the trivial solution of a circular bubble travelling with speed $U=2$, labelled $m=0$.  The nontrivial branches, $m=1,2,\ldots$, all have $U<2$.  We have applied techniques in exponential asymptotics to describe these solution branches in the limit $B\rightarrow 0$.  In particular, our analysis leads to the solvability condition (\ref{4:fullcriteria}), which is a relationship that provides an estimate for $U$ in terms of $B$ and $m$.  By comparing with numerical solutions, we have demonstrated that this estimate works very well for small to moderate values of surface tension.

The techniques in exponential asymptotics that we employ were developed by Chapman, King \& Adams~\cite{chapman98} and subsequently used in a variety of problems in applied mathematics.  Broadly speaking, the strategy is to a) derive a form for the late-order terms in an asymptotic power series by proposing a factorial-over-power ansatz which includes a singulant function $\chi$ on the denominator and a prefactor function $F$ on the numerator; b) locate potential Stokes lines by setting $\mathrm{Im}[\chi] = 0$, noting they emerge at singularities in the complex plane and intersect the boundary of the physical problem; and c) truncate the series optimally (at the least term) and determine how the exponentially small remainder switches on and off across Stokes lines.  In our case, we have two key remainder terms that switch on and off in a way that suggest they both become exponentially large in a particular region of the domain; we derive our solvability condition (\ref{4:fullcriteria}) by arguing that a physically meaningful solution is only possible if both of these contributions cancel each other out.  While exponential asymptotics has been used to study similar selection problems with bubbles and fingers in Hele-Shaw flow \cite{Combescot,CombescotEtAl,Hong,Shraiman,Tanveer86,Tanveer872,Tanveer89}, we believe the use of this method provide a more transparent analysis, providing clarity about where and how the exponentially small remainder terms come into play.

An important result of our work is the scaling law (\ref{eq:mainresult}), which comes from simplifying our solvability condition (\ref{4:fullcriteria}) for $B\ll 1$.  This result confirms the $B^2$ scaling predicted by \cite{GreenLustriMcCue,HongFamily88}, and goes further by giving the prefactor that explicitly provides the dependence on the solution branch $m$.  This scaling law is interesting since it is different to the analogous relationship for selection problems in a Hele-Shaw channel, namely $U\sim 2- k B^{2/3}$, where $k$ is function of $m$.  It is worth repeating that this difference is possible because the limits that the bubble area vanishes and $B\rightarrow 0$ do not commute for the channel problem \cite{GreenLustriMcCue,HongFamily88,Tanveer89}.

An interesting feature of our results is that the front of the bubble becomes nonconvex for sufficiently high surface tension values, with a certain number of dimples appearing, depending on the branch number.  For $m=1$ the bubbles have a double tip (first noted by Tanveer~\cite{Tanveer86}).  For $m=2$ there are three tips.  This pattern continues so that we have bubbles with $m+1$ tips on the $m$th solution branch, with the right-most point of symmetry alternating between being locally convex ($m$ even) to locally concave ($m$ odd) as $m$ increases.  We have been able to explain that the cause of this pattern is the oscillatory nature of the exponentially small remainder terms that arise after optimal truncation.  A perfectly analogous pattern appears for Saffman-Taylor fingers on increasing solution branches~\cite{GardinerEtAlb}, and we expect the same explanation to carry over to that well-studied problem.  As far as we are aware, our analysis provides the first theoretical explanation for these exotic-shaped interfaces.

It is worth commenting on the stability and time-dependent solutions.  There are linear stability results for small bubbles in a finite-width channel~\cite{tanveersaffman87} and rigorous results for a two-phase problem~\cite{YeTanveer11} which suggest the $m=0$ branch of circular solutions is stable, while the other branches are unstable.  On the other hand, time-dependent solutions may exhibit intermediate-time behaviour where, for example, the bubble boundary may evolve to a solution on the $m=1$ branch before ultimately either tending to the $m=0$ solution or breaking up into two separate bubbles.  This type of behaviour can occur in the analogous Saffman-Taylor finger problem, where stability results are studied in~\cite{kesslerlevine85,tanveer87_2318}, while the intermediate-time behaviour is discussed in the supplementary material of~\cite{GardinerEtAlb}.

More generally, double-tipped bubbles like those in the $m=1$ branch have been observed numerically by Meiburg~\cite{Meiburg89}, for example, by explicitly incorporating additional (three-dimensional) effects that are not included in the standard Hele-Shaw models.  Further, experimental results of Kopf-Sill \& Homsy~\cite{KopfSill88} demonstrate the existence of these double-tipped bubbles, although the quantitative agreement with the analytical predictions for the speed of the bubbles is not strong.  Finally, similar double-tipped (or ``dimpled'') bubbles have been observed very recently in an experimental Hele-Shaw cell with channel-depth perturbations by Franco-G\'omez et al.~\cite{FrancoGomez2018}.  Other multiple-tipped bubble shapes have be generated using this type of apparatus~\cite{deLozar2009,FrancoGomez2016,Hazel2013,Thompson2014}.  In all of these cases, there are important three-dimensional effects, and these effects are not considered in our mathematical model.  However, we expect that our description of the emergence of multiple-tipped bubble shapes via exponential small wave-like contributions in the limit of small surface tension to carry over to the more complicated and, perhaps, more physical realistic scenarios.

%Whole lot of other related work on bubbles translating in Hele-Shaw, including analytical and numerical results for zero-surface-tension models, both steady \cite{aldushin99} [CCG's papers and others] and time-dependent, numerical solutions with 3D effects [Zhu Gallaire, Ling16, Meiburg88], as well as wealth of older and more recent experimental studies \cite{keiser18,KopfSill88,maxworthy86,reichert18}.

\section*{Acknowledgments}
CCG and SWM acknowledge the support of the Australian Research Council Discovery Project DP140100933. CJL and CCG acknowledge the support of Macquarie University Research Seeding Grant 72585747. CCG acknowledges the support of Australian Research Council Discovery Early Career Researcher Award DE180101098.

\appendix
\section{Determining $\Lambda$}\label{App:Lambda}

In order to determine the quantity $\Lambda$, we must match the outer expansion \eqref{0:ansatz} with an inner expansion in the neighbourhood of $\zeta = \i/q$. The appropriate inner scaling occurs where the asymptotic series \eqref{0:fseries} to break down, as successive terms are no longer smaller than previous terms in the asymptotic limit.  From \eqref{0:innerofouter}, we see that the factorial-over-power ansatz breaks down near the singularity in a region of width $\zeta - \i/q = \mathcal{O}(B^{2/7})$ as $B \rightarrow 0$. We therefore consider the leading-order behaviour of a local expansion in the neighbourhood of the point $\zeta = \i/q$, and match this with the outer expansion. This matching process is discussed in detail in \cite{chapman98}, and we will provide an outline of the calculations here.

%
%\begin{figure}
%\centering
%\caption{Evaluating the ratio in \eqref{A:Lambda} with $U = 1.5$ for increasing values of $n$, up to $n = 50$. We note that this ratio converges rapidly, providing a precise approximation for $\Lambda$.}\label{F:Lambda}
%\end{figure}

%\begin{equation*}
%\zeta=\i/q+B^{2/7}\eta,\qquad g(\eta) = B^{3/2} f(\zeta).
%\end{equation*}

We define the new inner variables $\zeta=\i/q+B^{2/7}\eta$ and $g(\eta) = B^{3/2} f(\zeta)$. Simplifying the governing equation \eqref{0:analytic} and applying the inner expressions given in \eqref{0:localvals} gives the inner equation to leading order as $B \rightarrow 0$ as
\begin{align}
 \frac{16q^2 aU (U-1)}{(2-U)}  \Bigg(\frac{2\i a}{U^{3/2}} (2-U)^{3/2}&\eta + g'(\eta)\Bigg)^3g^2(\eta) + \Bigg(\frac{2\i a}{U^{3/2}}  (2-U)^{3/2}+ g''(\zeta)\Bigg)^2 = 0.\label{A:Innergov}
%\frac{16q^2 aU (U-1)}{(2-U)}  \Bigg(\frac{2\i a}{U^{3/2}} (2-U)^{3/2}&\eta + g'(\eta)\Bigg)^3g^2(\eta)\\
%& + \Bigg(\frac{2\i a}{U^{3/2}}  (2-U)^{3/2}+ g''(\zeta)\Bigg)^2 = 0.\label{A:Innergov}
\end{align}
We expand $g(\eta)$ as the local series
\begin{equation}
g(\eta) = \sum_{n=0}^{\infty} \frac{g_n}{\eta^{3/2 + 7r/2}}\label{A:innerexp}
\end{equation}
and substitute this expression into \eqref{A:Innergov} to recursively obtain $g_n$ for arbitrarily large values of $n$.

Matching the outer limit of the inner expansion \eqref{A:innerexp} with inner limit of the late-order outer expansion \eqref{0:innerofouter} gives
\begin{equation}
\Lambda C = \lim_{n \rightarrow \infty}\frac{g_n \alpha^{2n - 13/14}}{\Gamma(2n-13/14)}.\label{A:Lambda}
\end{equation}
While this limit cannot be evaluated analytically, we can compute $g_n$ for large values of $n$, and approximate $\Lambda$ numerically by evaluating \eqref{A:Lambda} for large finite values of $n$. For example, using $U = 1.5$ and $n = 50$, we find that $\Lambda \approx -0.079 + 0.347\i$. Changing the value of $U$ produces different values for $\Lambda$, however the process for computing this quantity remains the same.

%In Figure \ref{F:Lambda}, we demonstrate an example calculation for $U = 1.5$, showing the value of $\Lambda$ that would be obtained by using a range of $n$ values to evaluate the expression \eqref{A:Lambda}. Using $n = 50$, we find that $\Lambda \approx -0.079 + 0.347\i$. Changing the value of $U$ produces different values for $\Lambda$, however the process for computing this quantity remains the same.


\begin{thebibliography}{10}

\bibitem{abramstegun} {\sc M. Abramowitz and I.~A. Stegun}, {\em Handbook of Mathematical Functions with Formulas, Graphs, and Mathematical Tables}, New York, Dover.

\bibitem{aldushin99}
{\sc A.P. Aldushin and B.J. Matkowsky}, {\em Extremum princples for selection in the Saffman-Taylor finger and Taylor-Saffman bubble problems}, Phys. Fluids, (1999), pp.~1287--1296.

\bibitem{Alimov2016}
{\sc M.M. Alimov}, {\em Unsteady motion of a bubble in a Hele-Shaw cell}, Fluid Dynamics, 51 (2016), pp.~253--265.

\bibitem{BenJacobGarik90}
{\sc E. Ben-Jacob and P. Garik}, {\em The formation of patterns in non-equilibrium growth}, Nature, 343 (1990), pp.~523--530.

\bibitem{BerryHowls90}
{\sc M. V. Berry and C. J. Howls}, {\em Hyperasymptotics}, Proc. R. Soc. A, 430(1880):653-668, (1990).

\bibitem{BodyKing05}
{\sc G.L. Body, J.R. King and R.H. Tew}, {\em Exponential asymptotics of a fifth-order partial differential equation}, Euro. J. Appl. Math., 16 (2005), pp.~647-681.

\bibitem{Boyd99}
{\sc J. P. Boyd}, {\em The devil's invention: asymptotic, superasymptotic and hyperasymptotic series}, Acta Appl. Math., 56 (1999), pp.~1--98.

\bibitem{Casademunt04}
{\sc J. Casademunt}, {\em Viscous fingering as a paradigm of interfacial pattern formation: Recent results and new challenges}, Chaos, 14 (2004), pp.~809--824.

\bibitem{Chapman99}
{\sc S.J. Chapman}, {\em On the role of Stokes lines in the selection of Saffman-Taylor fingers with small surface tension}, Eur. J. Appl. Math., 10 (1999), pp.~513--534.

\bibitem{ChapmanKing}
{\sc S.J. Chapman, J.R. King}, The selection of Saffman-Taylor fingers by kinetic undercooling. {\em J. Eng. Math.}, 46 (2003), pp.~1--32.

\bibitem{chapman98}
{\sc S.J. Chapman, J.R. King and K.L. Adams}, {\em Exponential asymptotics and Stokes lines in nonlinear ordinary differential equations}, Proc. R. Soc. Lond. A, 454 (1998), pp.~2733--2755.

\bibitem{chapmanmortimer05}
{\sc S.J. Chapman and D.B. Mortimer}, {\em Exponential asymptotics and Stokes lines in a partial differential equation}, Proc. R. Soc. A, 461 (2005), pp.~2385--2421.

%\bibitem{chapmanVDB1}
%{\sc S.J. Chapman and J.-M. Vanden-Broeck}, {\em Exponential asymptotics and capillary waves}, SIAM J. Appl. Math., 62 (2002), pp.~1872--1898.

\bibitem{chapmanVDB2}
{\sc S.J. Chapman and J.-M. Vanden-Broeck}, {\em Exponential asymptotics and gravity waves}, J. Fluid Mech., 567 (2006), pp.~299--326.



\bibitem{Combescot}
{\sc R. Combescot and T. Dombre}, {\em Selection in the Saffman-Taylor bubble and asymmetrical finger problem}, Phys. Rev. A, 38 (1988), 2573--2581.

\bibitem{CombescotEtAl}
{\sc R. Combescot, T. Dombre, V. Hakim, Y. Pomeau and A. Pumir}, {\em Shape selection of Saffman-Taylor fingers}, Phys. Rev. Lett., 56 (1986), 2036.

\bibitem{Crowdy2009}
{\sc D.G. Crowdy.}, {\em Multiple steady bubbles in a Hele-Shaw cell}, Proc. R. Soc. A, 465 (2009), pp.~421--435.

%\bibitem{Dallaston2012}
%M.C. Dallaston, S.W. McCue.  New exact solutions for Hele-Shaw flow in doubly connected regions.  {\em Phys. Fluids}, 24 (2012), 052101.

%\bibitem{Dallaston}
%M.C. Dallaston, S.W. McCue. Corner and finger formation in Hele-Shaw flow with kinetic undercooling regularisation.. {\em Eur. J. Appl. Math.}, 25 (2014), pp.~707--727.

\bibitem{Dingle}
{\sc R. B. Dingle}. {\em Asymptotic expansions: their derivation and interpretation}. London, Academic Press, (1973).

\bibitem{deLozar2009} A. De L\'ozar, A. Heap, F. Box, A.L. Hazel, A. Juel.  Tube geometry can force switchlike transitions in the behavior of propagating bubbles.  {\em Phys. Fluids}, 21 (2009), 101702.

\bibitem{Dorsey}
{\sc A.T. Dorsey and O. Martin}, {\em Saffman-Taylor fingers with anisotropic surface tension}, Phys. Rev. A, 35 (1987), 3989(R).

\bibitem{FrancoGomez2016}
{\sc A. Franco-G\'omez, A.B. Thompson, A.L. Hazel and A. Juel}, {\em Sensitivity of Saffman-Taylor fingers to channel-depth perturbations}, J. Fluid Mech., 794 (2016), pp.~343--368.

\bibitem{FrancoGomez2018} {\sc A. Franco-G\'omez, A.B. Thompson, A.L. Hazel and A. Juel}, {\em Bubble propagation in Hele-Shaw channels with centred constrictions}, Fluid Dyn. Res., 50 (2018), 021403.

\bibitem{GardinerEtAlb}
{\sc B.P.J. Gardiner, S.W. McCue and T.J. Moroney}, {\em Discrete families of Saffman-Taylor fingers with exotic shapes}, Results in Physics, 5 (2015), pp.~103--104, .

%\bibitem{GardinerEtAl2} {\sc B.P.J. Gardiner, S.W. McCue, M.C. Dallaston and T.J. Moroney}, {\em Saffman-Taylor fingers with kinetic undercooling}, Phys. Rev. E, 91 (2015), 023016.

\bibitem{GreenLustriMcCue}
{\sc C.C. Green, C.J. Lustri and S.W. McCue}, {\em The effect of surface tension on steadily translating bubbles in an unbounded Hele-Shaw cell}, Proc. R. Soc. A., 473 (2017), 20170050.

%\bibitem{cohenerneuxII} {\sc D.~S. Cohen and T. Erneux}, {\em Free boundary problems in controlled release pharmaceuticals. II: swelling-controlled release}, SIAM J. Appl. Math., 48 (1988), pp.~1466--1474.

\bibitem{GreenVas14}
{\sc C.C. Green and G.L. Vasconcelos}, {\em Multiple steady bubbles in a Hele-Shaw channel}, Proc. R. Soc. A, 470 (2014), 20130698.


\bibitem{Hazel2013}
{\sc A.L. Hazel, M. Pailha, S.J. Cox and A. Juel}, {\em Multiple states of finger propagation in partially occluded tubes}, Phys. Fluids, 25 (2013), 062106.

\bibitem{HongFamily88}
{\sc D.C. Hong and F. Family}, {\em Bubbles in the Hele-Shaw cell: pattern selection and tip perturbations}, Phys. Rev. A, 38 (1988), pp.~5253--5259.

\bibitem{Hong}
{\sc D.C. Hong and J.S. Langer}, {\em Analytic theory of the selection mechanism in the Saffman-Taylor problem}, Phys. Rev. Lett., 56 (1986), 2032.

\bibitem{Homsy87}
{\sc G.M. Homsy}, {\em Viscous fingering in porous media}, Ann. Rev. Fluid Mech., 19 (1987), pp.~271--311.

\bibitem{kessler88}
{\sc D.A. Kessler, J. Koplik and H. Levine}, {\em Pattern selection in fingered growth phenomena}, Adv. Phys., 37 (1988), pp.~255--339.

\bibitem{kesslerlevine85}
{\sc D.A. Kessler and H. Levine}, {\em Stability of finger patterns in Hele-Shaw cells}, Phys. Rev. A, 32 (1985), pp.~1930--1933.


\bibitem{keiser18}
{\sc L. Keiser, K. Jaafar, J. Bico and E. Reyssat}, {\em Dynamics of non-wetting drops confined in a Hele-Shaw cell}, J. Fluid Mech., 845 (2018), pp.~245--262.


\bibitem{Khalid2015}
{\sc A.H. Khalid, N.R. McDonald and J.-M. Vanden-Broeck}, {\em On the motion of unsteady translating bubbles in an unbounded Hele-Shaw cell}, Phys. Fluids, 27 (2015), 012102.

\bibitem{kingchapman01}
{\sc J.R. King and S.J. Chapman}, {\em Asymptotics beyond all orders and Stokes lines in nonlinear differential-difference equations}, Euro. J. Appl. Math., 12 (2001), pp.~433--463.

\bibitem{KopfSill88}
{\sc A.R. Kopf-Sill and G.M. Homsy}, {\em Bubble motion in a Hele-Shaw cell}, Phys. Fluids, 31 (1988), pp.~18--26.

\bibitem{Langer89}
{\sc J.S. Langer}, {\em Dendrites, viscous fingers, and the theory of pattern formation}, Science, 243 (1989), pp.~1150--1156.

\bibitem{Ling16}
{\sc Y. Ling, J.-M. Fullana, S. Popinet and C. Josserand}, {\em Droplet migration in a Hele-Shaw cell: effect of the lubrication film on the droplet dynamics}, Phys. Fluids, 28 (2016), 062001.

\bibitem{lustrichapman13}
{\sc C.J. Lustri and S.J. Chapman}, {\em Steady gravity waves due to a submerged source}, J. Fluid Mech., 732 (2013), pp.~660--686.

\bibitem{lustrichapman14}
{\sc C.J. Lustri and S.J. Chapman}, {\em Unsteady flow over a submerged source with low Froude number}, Euro. J. Appl. Math., 25 (2014), pp.~655--680.

\bibitem{lustrimccue12}
{\sc C.J. Lustri, S.W. McCue and B.J. Binder}, {\em Free surface flow past topography: a beyond-all-orders approach}, Euro. J. Appl. Math., 23 (2012), pp.~441--467.

\bibitem{lustrimccue13}
{\sc C.J. Lustri, S.W. McCue and S.J. Chapman}, {\em Exponential asymptotics of free surface flow due to a line source}, IMA J. Appl. Math., 78 (2013), pp.~697--713.

\bibitem{maxworthy86}
{\sc T. Maxworthy}, {\em Bubble formation, motion and interaction in a Hele-Shaw cell}, J. Fluid Mech., 173 (1986), pp.~95--114.

\bibitem{Meiburg89}
{\sc E. Meiburg}, {\em Bubbles in a Hele-Shaw cell: Numerical simulation of three-dimensional effects}, Phys. Fluids A, 1 (1989), pp.~938--946.

\bibitem{McLeanSaffman}
{\sc J.W. McLean and P.G. Saffman}, {\em The effect of surface tension on the shape of fingers in a Hele-Shaw cell}, J. Fluid Mech., 102 (1981),  pp.~455--469.

\bibitem{Mineev}
{\sc M. Mineev-Weinstein}, {\em Selection of the Saffman-Taylor finger width in the absence of surface tension: an exact result},  Phys. Rev. Lett., 80 (1998), 2113.

\bibitem{reichert18}
{\sc B. Reichert, A. Huerre, O. Theodoly, M.-P. Valignat, I. Cantat and M.-C. Jullien}, {\em Topography of the lubrication film under a
pancake droplet travelling in a Hele-Shaw cell}, J. Fluid Mech., 850 (2018), pp.~708--732.

\bibitem{ST}
{\sc P.G. Saffman and G.I. Taylor}, {\em The penetration of a fluid into a porous medium or Hele-Shaw cell containing a more viscous liquid}, Proc. R. Soc. A, 245 (1958), pp.~312--329.

\bibitem{Seguretal91}
{\sc H. Segur, S. Tanveer and H.J. Levin}, {\em Asymptotics Beyond All Orders}, 1991, Springer US, New York.

\bibitem{Shraiman}
{\sc B.I. Shraiman}, {\em Velocity selection and the Saffman-Taylor problem}, Phys. Rev. Lett., 56 (1986), 2028.

\bibitem{Tanveer86}
{\sc S. Tanveer}, {\em The effect of surface tension on the shape of a Hele-Shaw cell bubble}, Phys. Fluids, 29 (1986), pp.~3537--3548.

\bibitem{Tanveer87}
{\sc S. Tanveer}, {\em New solutions for steady bubbles in a Hele-Shaw cell}, Phys. Fluids, 30 (1987), pp.~651--658.

\bibitem{tanveer87_2318}
{\sc S. Tanveer}, {\em Analytic theory for the linear stability of the Saffman-Taylor finger}, Phys. Fluids 30 (1987), pp.~2318--2329.

\bibitem{Tanveer872}
{\sc S. Tanveer}, {\em Analytic theory for the selection of a symmetric Saffman-Taylor finger in a Hele-Shaw cell}, Phys. Fluids 30 (1987), pp.~1589--1605.

\bibitem{Tanveer89}
{\sc S. Tanveer}, {\em Analytic theory for the determination of velocity and stability of bubbles in a Hele-Shaw cell, part I: velocity selection}, Theoret. Comput. Fluid Dynamics, 1 (1987), pp.~135--163.

\bibitem{tanveersaffman87}
{\sc S. Tanveer and P.G. Saffman}, {\em Stability of bubbles in a Hele-Shaw cell}, Phys. Fluids, 30 (1987), pp.~2624--2635.

\bibitem{TS}
{\sc G.I. Taylor and P.G. Saffman}, {\em A note on the motion of bubbles in a Hele-Shaw cell and porous medium}, Q. J. Mech. Appl. Math., 12 (1959), pp.~265--279.

\bibitem{Thompson2014}
{\sc A.B. Thompson, A. Juel and A.L. Hazel}, {\em Multiple finger propagation modes in Hele-Shaw channels of variable depth},  J. Fluid Mech., 746 (2014), 1pp.~23--164.

\bibitem{trinhchapman13}
{\sc P.H. Trinh and S.J. Chapman}, {\em New gravity–capillary waves at low speeds. Part 1. Linear geometries}, J. Fluid Mech., 724 (2013), pp.~367--391.

\bibitem{trinhchapman14}
{\sc P.H. Trinh and S.J. Chapman}, {\em The wake of a two-dimensional ship in the low-speed limit: results for multi-cornered hulls}, J. Fluid Mech., 741 (2014), pp.~492--513.

\bibitem{VDB1983}
{\sc J.-M. Vanden-Broeck}, {\em Fingers in a Hele-Shaw cell with surface tension}, Phys. Fluids, 26 (1983), 2033.

%\bibitem{Vasconcelos2015}
%G.L. Vasconcelos. Multiple bubbles and fingers in a Hele-Shaw channel: complete set of steady solutions.  {\em J. Fluid Mech.}, 780 (2015),  pp.~299--326.

\bibitem{VMW2014}
{\sc G.L. Vasconcelos and M. Mineev-Weinstein}, {\em Selection of the Taylor-Saffman bubble does not require surface tension}, Phys. Rev. E, 89 (2014), 061003(R).

\bibitem{YeTanveer11}
{\sc J. Ye and S. Tanveer}, {\em Global existence for a translating near-circular Hele-Shaw bubble with surface tension}, SIAM J. Math. Anal., 43 (2011), pp.~457--506.


%\bibitem{cohenerneuxII} {\sc D.~S. Cohen and T. Erneux}, {\em Free boundary problems in controlled release pharmaceuticals. II: swelling-controlled release}, SIAM J. Appl. Math., 48 (1988), pp.~1466--1474.

\end{thebibliography}
\end{document}